\documentclass[twocolumn]{aastex631}

\usepackage{amsmath}

\graphicspath{{./}}

\shorttitle{LAEs with multiple components}
\shortauthors{Liu et al.}

\begin{document}

\title{A high rate of foreground contaminants toward high-redshift galaxies resolved by JWST}

\author[0000-0002-4385-0270]{Weiyang Liu}
\affiliation{Department of Astronomy, School of Physics, Peking University, Beijing 100871, China}
\affiliation{Kavli Institute for Astronomy and Astrophysics, Peking University, Beijing 100871, China}

\author[0000-0003-4176-6486]{Linhua Jiang}
\affiliation{Department of Astronomy, School of Physics, Peking University, Beijing 100871, China}
\affiliation{Kavli Institute for Astronomy and Astrophysics, Peking University, Beijing 100871, China}

\begin{abstract}
We present a study of high-redshift Ly$\alpha$ emitters (LAEs) with multiple components using HST and JWST. High-redshift galaxies are mostly point-like objects on ground-based images, but they often exhibit multiple components in higher spatial resolution images. JWST for the first time allow detailed analyses on these individual components. We collect 840 spectroscopically confirmed LAEs at $z=2\sim7$ from the literature and nearly 50\% of them appear to have multiple components in JWST images. We further construct a sample of 248 LAEs that have two or more relatively isolated components in a circular aperture of 2$\arcsec$ in diameter. We estimate photometric redshifts for all 593 components of the 248 LAEs, and find that 68\% of them are `real components' with photometric redshifts consistent with the spectroscopic redshifts of the LAEs. The remaining components are mostly foreground objects. The fraction of the `real components' decreases rapidly with the projected distance to the LAE centers from $\sim80\%$ at $0\farcs2-0\farcs4$ to $\sim30\%$ at $0\farcs8-1\farcs0$. Our SED modeling results suggest that the majority of the LAEs are young, low-mass, low extinction starburst galaxies (partly due to a selection effect), and their `real components' have stronger star-forming activities than main-sequence galaxies. We investigate the potential impact of the high foreground contamination rate on previous studies based on ground-based images that often use a 2$\arcsec$ aperture for photometry, and find that some of key parameters such as stellar mass would have been largely affected.

\end{abstract}


\section{Introduction} \label{sec:intro}

Galaxies with strong Ly$\alpha$ emission, or Ly$\alpha$ emitters (LAEs), were theoretically predicted more than half century ago by \cite{1967ApJ...147..868P} and first discovered observationally in the 1990s \citep{1996Natur.380..411P, 1996Natur.382..231H, 1996Natur.383...45P}. They are powerful tools for studying galaxy formation and evolution, the epoch of reionization (EoR), and cosmology \citep[e.g.,][]{2020ARA&A..58..617O, 2016ARA&A..54..761S}. LAEs are usually regarded as young, low-mass, low dust-extinction, star-forming galaxies, representing the early phase of galaxy evolution \citep[e.g.,][]{2024ApJ...963...97I, 2025arXiv250108568F}. Due to the strong Ly$\alpha$ line emission at $1215.67\rm \, \AA$ in the rest frame, it is relatively easy to spectroscopically identify high-redshift LAEs, including those at the EoR that allow us to study EoR through the absorption of Ly$\alpha$ photons by neutral hydrogen gas in the intergalactic medium (IGM) \citep[e.g.,][]{2022ApJ...926..230N, 2024A&A...683A.238J, 2024ApJ...967...28N}. In addition, large LAE samples with robust spectroscopic redshifts can be used to sample the matter distribution at high redshift and thus constrain cosmological parameters \citep[e.g.,][]{2021ApJ...923..217G}.

Several methods have been used to find LAEs. One can take spectra of color-selected Lyman-break galaxies (LBGs) to find those with strong Ly$\alpha$ emission lines, but the efficiency is relatively low \citep{2009ApJ...695.1163V, 2018A&A...619A.147P}. A popular and more efficient method is the narrowband technique, which selects LAE candidates using a narrowband filter centered on the redshifted Ly$\alpha$ emission line and identifies LAEs through follow-up spectroscopy \citep[e.g.,][]{2008ApJS..176..301O, 2017ApJ...846..134J}. In recent years, integral field units (IFUs) on large telescopes have performed blind LAE surveys without pre-selection and have successfully found a large number of LAEs, including many very faint LAEs due to their higher spectral resolutions compared with narrowband filters \citep{2017A&A...608A...2I, 2019A&A...624A.141U, 2023ApJ...943..177M}. Slitless spectra from space telescopes such as HST found LAEs at relatively lower redshift owing to their UV capability \citep{2011ApJ...738..136C, 2014ApJ...783..119W} and will be a promising method to find high-redshift LAEs with future space telescopes.

Most LAEs show compact morphology with a single component in ground-based images. In higher spatial resolution images from space telescopes such as HST, some of these single-component LAEs show clumpy morphology with two or more components  \citep[e.g.,][]{2016ApJ...819...25K}. Such multi-component LAEs could potentially affect the results of ground-based LAE studies such as the Ly$\alpha$ luminosity function (LF), the Ly$\alpha$ equivalent width (EW), and the physical properties of LAEs, since we do not know which components are actually emitting Ly$\alpha$ photons. Therefore, it is crucial to figure out whether these components are at the same redshift as the spectroscopic redshifts (spec-$z$) of the LAEs, or they are just foreground/background interlopers. In addition, it is also important to understand the physical properties of these components (if they are at the same redshift) and possible relationship between them.

The James Webb Space Telescope \citep[JWST;][]{2023PASP..135f8001G} has high spatial resolution in the near-IR, and for the first time provides detailed information on multiple components of LAEs in the rest-frame optical range. This is crucial for estimating their photometric redshifts (photo-$z$) and derive their physical properties. In this paper, we investigate LAEs with multiple components in a circular aperture of $2\arcsec$ diameter using high-angular-resolution HST and JWST images. We use $2\arcsec$ because it is a typical size used in ground-based narrowband LAE photometry \citep[e.g.,][]{2008ApJS..176..301O, 2017ApJ...846..134J}. Section \ref{sec:imgDatPhot} describes our data reduction and photometry. In Section \ref{sec:LAEsample} we present our LAE sample. In Section \ref{sec:Photoz} we show the photo-$z$ results and the fraction of components at the same redshift as the spec-$z$ of LAEs. In Section \ref{sec:SED} we show the results of the spectral energy distribution (SED) fitting and LAE properties. We discuss how multi-component features affect the results of ground-based LAE studies in Section \ref{sec:discussion}. We give a summary in Section \ref{sec:summary}. Throughout this paper, all magnitudes are in the AB system. We adopt a $\Lambda$-dominated flat cosmology with $H_0=70 \rm \, km \, s^{-1} \, Mpc^{-1}$, $\Omega_m=0.3$, and $\Omega_\Lambda=0.7$.

\section{Data and data reduction} \label{sec:imgDatPhot}

In this section, we describe the HST and JWST imaging data used in this paper, the reduction procedure of the JWST images, and the photometry of our targets in the images.

\subsection{HST images}

The LAEs in this paper are selected in four deep fields, GOODS-S, GOODS-N, COSMOS, and UDS. There are many reduced HST images in these fields and we directly use them in this study. For the GOODS-S and GOODS-N fields, we use image products in the F435W, F606W, F775W, F814W, and F850LP bands from the Hubble Legacy Fields (HLF) \citep{2016arXiv160600841I, 2019ApJS..244...16W}. For the COSMOS and UDS fields, we use image products in the F606W and F814W bands from the HST CANDELS survey \citep{2011ApJS..197...35G, 2011ApJS..197...36K}. The pixel scales of these images are $0\farcs03$. The depths of the HST images are listed in Table \ref{tab:limitmag}.

\begin{deluxetable}{ccccc}
\tablewidth{0pt}
\tablecaption{5$\sigma$ limiting magnitudes in a circular aperture of $0\farcs3$ diameter in the four fields. \label{tab:limitmag}}
\tablehead{
\colhead{Filter} & \colhead{GOODS-S} & \colhead{GOODS-N} & \colhead{COSMOS} & \colhead{UDS}  
}
\startdata
F435W & 28.79 & 28.78 & ... & ... \\
F606W & 29.11 & 28.79 & 28.06 & 28.17 \\
F775W & 28.54 & 28.58 & ... & ... \\
F814W & 28.79 & 28.81 & 27.86 & 28.12 \\
F850LP & 28.27 & 28.13 & ... & ... \\
F070W & 28.59 & 28.62 & ... & ... \\
F090W & 29.11 & 28.74 & 28.01 & 27.89 \\
F115W & 29.09 & 28.72 & 27.47 & 27.96 \\
F140M & 27.88 & ... & ... & ... \\
F150W & 29.09 & 28.80 & 27.73 & 28.16 \\
F162M & 29.79 & ... & ... & ... \\
F182M & 28.67 & 28.54 & ... & ... \\
F200W & 28.90 & 28.94 & 28.72 & 28.23 \\
F210M & 28.47 & 28.27 & ... & ... \\
F250M & 28.85 & ... & ... & ... \\
F277W & 29.55 & 29.48 & 28.53 & 28.76 \\
F300M & 30.24 & ... & ... & ... \\
F335M & 29.66 & 29.13  & ... & ... \\
F356W & 29.60 & 29.41 & 29.07 & 28.82 \\
F410M & 29.43 & 28.99 & 28.33 & 28.07 \\
F430M & 28.90 & ... & ... & ... \\
F444W & 29.30 & 29.11 & 28.38 & 28.51 \\
F460M & 28.68 & ... & ... & ... \\
F480M & 28.95 & ... & ... & ... 
\enddata
\end{deluxetable}

\subsection{JWST images and image reduction}

\begin{figure*}[t]
\epsscale{1.18}
\plotone{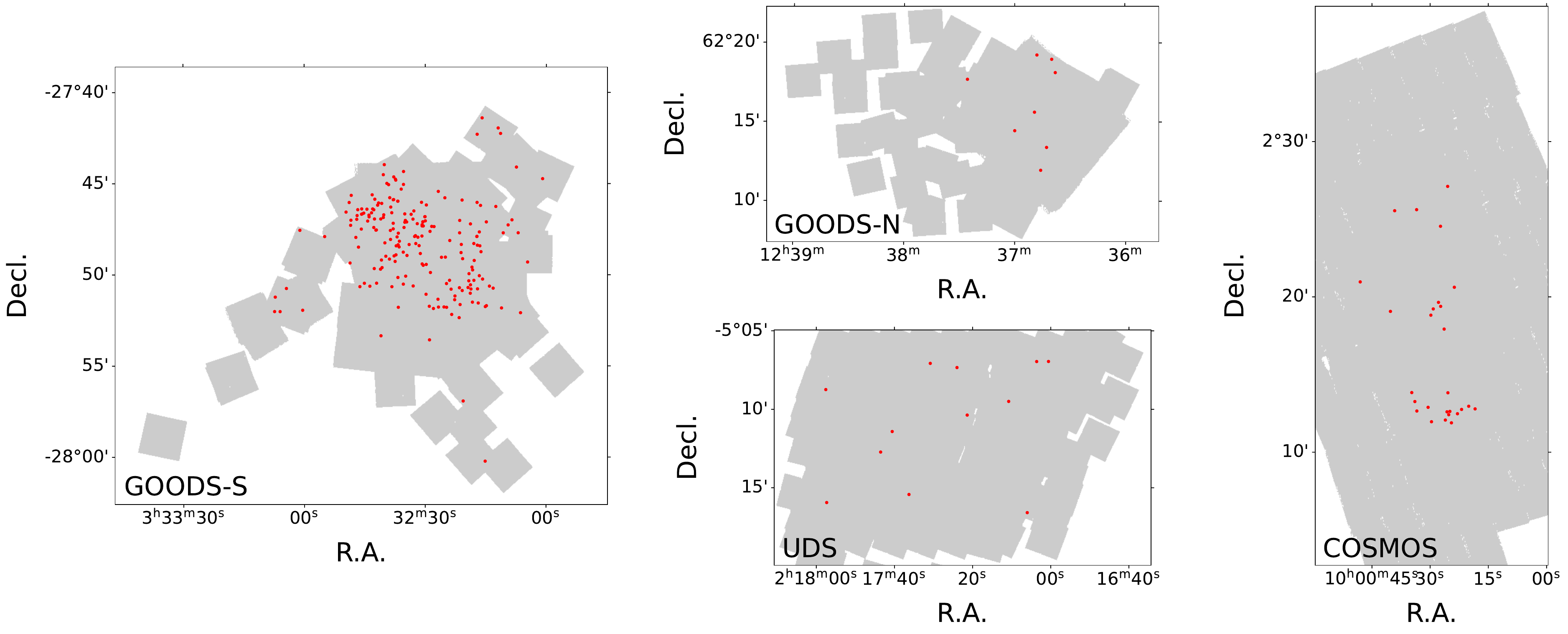}
\caption{Footprints of our JWST images. The gray area in the four panels show the footprints of our JWST images in the GOODS-S, GOODS-N, COSMOS, and UDS fields. A large part of them overlap with the HST HLF/CANDELS images. The red dots show the positions of all spectroscopically confirmed LAEs with multiple components in our LAE sample.
\label{fig:ImgFp}}
\end{figure*}

We download JWST NIRCam raw images {\tt\string uncal.fits} files in the GOODS-S, GOODS-N, COSMOS, and UDS fields from the MAST archive. The images in the GOODS-S field are from proposal IDs 1180, 1210, 1283, 1286, 1895, 1963, 2079, 2198, 2514, 2516, 3215, 3990, and 6541. The images in the GOODS-N field are from proposal IDs 1181, 1895, 2514, 2674, and 3577. The images in the COSMOS field are from proposal IDs 1635, 1727, 1810, 1837, 1840, 1933, 2321, 2362, 2514, 3990, and 6585. The images in the UDS field are from proposal IDs 1837, 1840, 2514, and 3990. All the JWST data used in this paper can be found in MAST:\dataset[https://doi.org/10.17909/fppb-3m14]{https://doi.org/10.17909/fppb-3m14}. We reduce raw data using the combination of the JWST Calibration Pipeline (v.1.12.5), some scripts from the CEERS NIRCam imaging reduction\footnote{https://github.com/ceers/ceers-nircam}, and our own custom scripts. The details of the CEERS NIRCam imaging reduction are presented in \cite{2023ApJ...946L..12B}. Here we give a brief description of the steps of our image reduction.

Stage 1 of the JWST Calibration Pipeline performs detector-level corrections, starting from {\tt\string uncal.fits} files and ending with {\tt\string rate.fits} files in units of count/s. Before running the ramp-fitting step in Stage 1 of the pipeline, we identify snowballs, enlarge their footprints, and flag them as {\tt\string JUMP\_DET} so that snowballs are reliably removed after the ramp-fitting step. The `Wisp' features from detector B4 of the F150W, F200W, and F210M images are subtracted using the wisp templates provided by the NIRCam team\footnote{https://stsci.box.com/s/1bymvf1lkrqbdn9rnkluzqk30e8o2bne}. The horizontal and vertical striping patterns in the images, i.e., 1/f noise, are subtracted with {\tt\string remstriping.py} in the CEERS team's scripts.

Then we run Stage 2 of the JWST Calibration Pipeline with the default parameters, which involves individual image calibrations such as flat-fielding and flux calibration. The output files are {\tt\string cal.fits}. We run Stage 3 of the JWST Calibration Pipeline to obtain a single mosaic for each filter in each field. Astrometry is calibrated using {\tt\string TweakregStep} with {\tt\string abs\_refcat} set to user-provided reference catalogs in the corresponding fields. We generate a reference catalog for GOODS-S/GOODS-N/COSMOS/UDS as follows. 
We first align all long wavelength (LW) F277W, F356W, F410M, and F444W images to detected objects in the HLF GOODS-S/HLF GOODS-N/CANDELS COSMOS/CANDELS UDS F814W image. We then combine all these LW images to make a mosaic image, and finally detect objects in the mosaic. After running {\tt\string SkyMatchStep} and {\tt\string OutlierDetectionStep}, we subtract a 2D background for each image following the method described in \cite{2023ApJ...946L..12B}, and then run {\tt\string ResampleStep} to drizzle and combine images to make one mosaic per filter. We set {\tt\string pixfrac} to be 0.8 and set pixel scale to be $0\farcs03$ for all filters. The {\tt\string output\_shape}, {\tt\string CRVAL} and {\tt\string CRPIX} values are the same for each field, so in each field the mosaics of all filters have the same WCS grid. Finally, a 2D background is subtracted from each mosaic using the method described in \cite{2023ApJ...946L..12B}. Figure \ref{fig:ImgFp} shows the footprints of our JWST mosaics in the GOODS-S, GOODS-N, COSMOS, and UDS fields. Note that the footprint sizes in some filters are much smaller than those shown in Figure \ref{fig:ImgFp}. The filters used in this paper and the depths of the JWST images are listed in Table \ref{tab:limitmag}.

\subsection{Photometry in the fields} \label{sec:photometry}

After obtaining these HST and JWST mosaic images, we do photometry for all objects in the four fields. Point spread functions (PSFs) in different filters are slightly different, so we do PSF homogenization before we do photometry in each filter. We first use the code {\tt\string SExtractor} \citep{1996A&AS..117..393B} to detect objects in each filter and plot them on the {\tt\string MAG\_AUTO} - full width half maximum (FWHM) diagram. We select point sources based on this diagram by requiring that the FWHM is consistent with a point source and the flux is not saturated. We visually inspect individual point sources and stack them to generate an empirical PSF using the function {\tt\string photutils.psf.EPSFBulider} in the package {\tt\string photutils} \citep{larry_bradley_2024_13989456}. The number of good point sources in each filter in each field ranges from 15 to 50. Second, we use the code {\tt\string PYPHER} \citep{2016A&A...596A..63B} to calculate a homogenization kernel between each filter and the reference filter F444W. These kernels are then convolved with the images to get PSF-homogenized images in all filters. We obtain the final detection image in each field by combining PSF-homogenized images in the F200W, F277W, F356W, F410M, and F444W bands (weighted by inverse variance). Third, we use the dual mode in {\tt\string SExtractor} to detect objects in the detection image with {\tt\string DETECT\_MINAREA} = 5, {\tt\string DETECT\_THRESH} = 1.3, {\tt\string DEBLEND\_NTHRESH} = 32, {\tt\string DEBLEND\_MINCONT} = 0.0001 and a top-hat convolution kernel with diameter = 4.0 pixels, and measure their flux in the PSF-homogenized images with {\tt\string PHOT\_AUTOPARAMS} = (1.1, 1.6) in all filters. The detection catalogs in the four fields are used in Section \ref{sec:LAEsample} to construct our LAE sample.

To measure the total flux of the objects, we do aperture corrections in two steps. The first step is to correct the flux in a small aperture to the flux in a larger aperture by calculating the flux ratio {\tt\string FLUX\_AUTO} (2.5, 3.5) / {\tt\string FLUX\_AUTO} (1.1, 1.6) in F444W as a function of {\tt\string MAG\_AUTO} (1.1, 1.6). The second step is to correct the aperture flux to the total flux with simulations. We simulate 3000 mock galaxies using the code {\tt\string GALFIT} \citep{2010AJ....139.2097P} with magnitudes following a uniform distribution between 23 and 27 mag, S\'{e}rsic index $n = 1.5$, and half-light radii following a log-normal distribution with a peak at 3.5 pixels ($\sim0.7$ kpc at $z\sim4.5$). These are typical values for high redshift LBGs \citep[e.g.,][]{2015ApJS..219...15S}. We then convolve them with the PSF of the F444W band. These mock galaxies are randomly put in blank regions in the PSF-homogenized images in the F200W, F277W, F356W, F410M, and F444W bands. We get a new detection image by combining these 5 images, and use the dual mode of {\tt\string SExtractor} to detect objects in the new detection image and measure flux in the simulated-galaxies-added F444W image with {\tt\string PHOT\_AUTOPARAMS} = (2.5, 3.5). We calculate the flux ratio of `simulated flux' / {\tt\string FLUX\_AUTO} (2.5, 3.5) in F444W as a function of {\tt\string MAG\_AUTO} (2.5, 3.5). By combining the two ratios, we correct the aperture flux of an object to its total flux. 

We estimate an independent flux error based on $\sigma_N = \sigma_1 \times \alpha \times N^\beta$ \citep{2023ApJ...946L..13F}, where $N$ is the number of pixels in an aperture, $\sigma_N$ is the flux error in an aperture of $N$ pixels that are not affected by the object flux, $\sigma_1$ is the standard deviation of the flux in all pixels not affected by the object flux, and $\alpha$ and $\beta$ are parameters to be fitted. The $\sigma_1$ value is obtained by calculating the sigma-clipped standard deviation in the blank regions of the original image. In order to get $\sigma_N$ with different pixel number $N$, we randomly select 5000 (500) circles with diameters of $1.5$ ($3.0$) arcsec in blank regions in each mosaic, and perform forced photometry in circular apertures with diameters ranging from 0.1 to 1.5 (1.6 to 3.0) arcsec concentric with these randomly selected circles. We then calculate the median absolute deviation (MAD) of the flux in these 5000 (500) circular apertures and multiply the MAD by 1.48 to calculate $\sigma_N$ at different $N$. After we have $\sigma_1$ and $\sigma_N$ for different aperture areas $N$, we fit $\alpha$ and $\beta$. Finally, we calculate the area $N$ of each detected object as $N = \pi$ $\times$ A\_IMAGE $\times$ B\_IMAGE $\times$ KRON\_RADIUS$^2$ and estimate its flux error with $\sigma_N = \sigma_1 \times \alpha \times N^\beta$.

\section{Results} \label{sec:result}

In this section, we first construct a sample of LAEs with multiple components. Then, we present the results of photo-$z$ and the physical properties of these components.

\subsection{LAE sample} \label{sec:LAEsample}

With the high resolution HST and JWST images in the four deep fields, we construct a sample of LAEs with multiple components. We first construct a parent LAE sample by compiling spectroscopically confirmed LAEs at $z\sim2-7$ in the literature covered by the JWST images. The LAEs in the parent sample were discovered by several different methods as described in Section \ref{sec:intro}. We include narrowband-selected and spectroscopically confirmed LAEs at redshift $z\sim2.1$ \citep{2014ApJ...788...74S, 2021MNRAS.505.1382M}, $z\sim2.8$ \citep{2016ApJS..226...23Z}, $z\sim3.1$ \citep{2008ApJS..176..301O, 2014MNRAS.439..446M, 2020ApJ...902..137G}, $z\sim3.7$ \citep{2008ApJS..176..301O, 2023ApJ...958..187L}, $z\sim4.5$ \citep{2013MNRAS.431.3589Z}, $z\sim5.7$ \citep{2008ApJS..176..301O, 2010ApJ...725..394H, 2012ApJ...760..128M, 2017ApJ...846..134J, 2020ApJ...903....4N}, $z\sim6.5$ \citep{2010ApJ...723..869O, 2010ApJ...725..394H, 2017ApJ...846..134J, 2022ApJ...926..230N}, $z\sim7$ \citep{2021NatAs...5..485H, 2022ApJ...934..167H}, and from the SILVERRUSH survey \citep{2023ApJS..268...24K}. In typical narrowband studies, the narrowband color excess roughly corresponds to rest-frame Ly$\alpha$ EW $\gtrsim 20 \rm \AA$ at $z<6$ \citep[e.g.,][]{2008ApJS..176..301O} or rest-frame Ly$\alpha$ EW $\gtrsim 10 \rm \AA$ at $z>6$ \citep[e.g.,][]{2010ApJ...723..869O} considering the IGM absorption in the EoR. In addition, ground-based narrowband studies usually find luminous LAEs with Ly$\alpha$ luminosities $\gtrsim 10^{42.2} \,\rm erg\, s^{-1}$ due to their limited depth of the narrowband images. In order to construct a uniform sample, we require that LAEs discovered by other methods should also meet the Ly$\alpha$ EW and Ly$\alpha$ luminosity criteria above. For example, we include LAEs discovered by spectroscopic follow-up observations of LBGs \citep[e.g.,][]{2009ApJ...695.1163V, 2012MNRAS.422.1425C, 2015A&A...576A..79L, 2017A&A...600A.110T, 2018A&A...619A.147P, 2020ApJ...904..144J, 2020ApJS..247...61F, 2023A&A...678A..25T} that meet the Ly$\alpha$ EW criterion. LAEs discovered with IFUs \citep{2017A&A...608A...2I, 2019A&A...624A.141U, 2022A&A...659A.183K} and other methods \citep[e.g.,][]{2018ApJ...858...94L, 2021MNRAS.503.4105T} that meet the Ly$\alpha$ EW and Ly$\alpha$ luminosity criteria are also included in our parent LAE sample. 

With the above steps, we obtain 989 LAEs, including 648 in GOODS-S, 20 in GOODS-N, 183 in COSMOS, and 138 in UDS. We further select LAEs covered by at least two HST bands and by at least six JWST bands. This is to ensure reliable measurements of LAE photo-$z$ and physical properties in the following analyses. This step results in a total of 840 LAEs, including 598 in GOODS-S, 17 in GOODS-N, 137 in COSMOS, and 88 in UDS.

Since our goal is to study LAEs with two or more components, we match the above LAE sample with the detection catalogs in Section \ref{sec:photometry} within a radius of 1$\arcsec$, and select LAEs that have two or more matches. Among the 840 LAEs in the sample, 408 LAEs satisfy this criterion, including 319 in GOODS-S, 11 in GOODS-N, 49 in COSMOS, and 29 in UDS. Another 366 LAEs have one match in the detection catalogs, including 229 in GOODS-S, 4 in GOODS-N, 77 in COSMOS, and 56 in UDS. The remaining 66 LAEs do not have any matched objects that have reliable flux measurement. The reasons include the large uncertainties of the LAE coordinates from ground-based observations and contamination by nearby bright objects or by diffraction spikes of bright stars. 

\begin{figure*}[t]
\epsscale{1.18}
\plotone{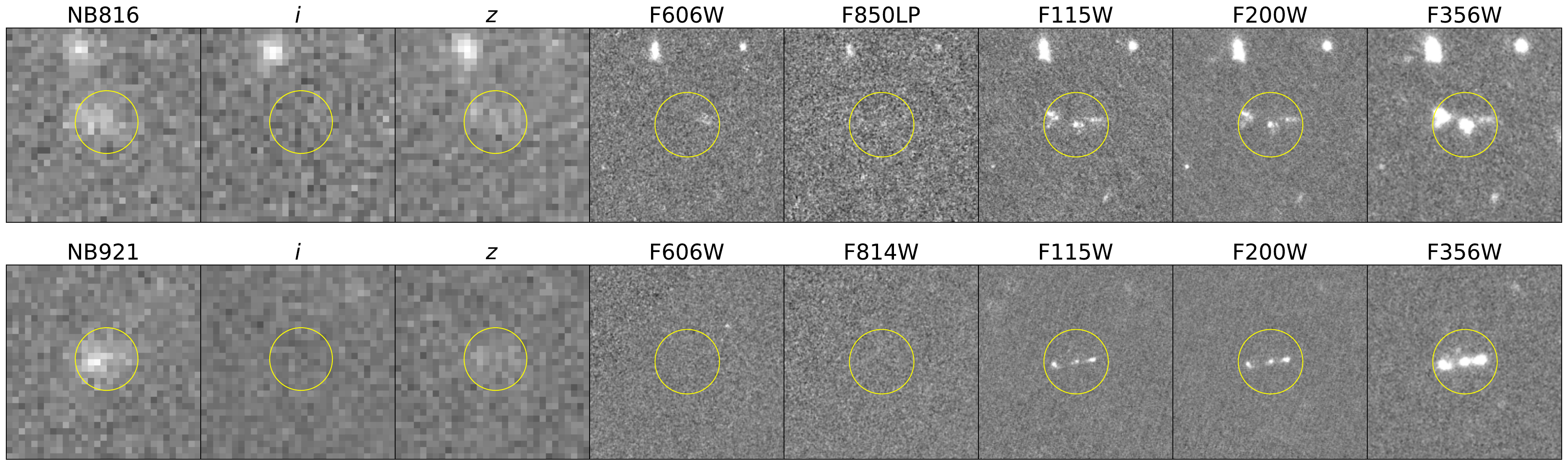}
\caption{Example of two multi-component LAEs in our LAE sample. The upper row shows a $z=5.656$ LAE from \cite{2017ApJ...846..134J} and the lower row shows a $z=6.595$ LAE from \cite{2010ApJ...723..869O}. The left three panels in each row show the ground-based narrowband, $i$, and $z$ band images. The right five panels show HST F606W, HST F850LP/F814W, JWST F115W, JWST F200W, and JWST F356W images. The sizes are $6\arcsec \times 6\arcsec$. The diameters of yellow circles are $2\arcsec$. 
\label{fig:LaeEx}}
\end{figure*}

\begin{deluxetable*}{cccccc}[t]
\tabletypesize{\footnotesize}
\tablewidth{0pt}
\tablecaption{Numbers of LAEs and components. \label{tab:numLAEComp}}
\tablehead{
\colhead{} & \colhead{GOODS-S} & \colhead{GOODS-N} & \colhead{COSMOS} & \colhead{UDS} & \colhead{Total}
}
\startdata
Number of LAEs & 201 & 8 & 27 & 12 & 248\\
Number of components & 487 & 18 & 61 & 27 & 593\\
Number of `real components' & 320 & 13 & 46 & 24 & 403\\
Number of foreground interlopers & 150 & 5 & 15 & 2 & 172\\
Number of background interlopers & 17 & 0 & 0 & 1 & 18\\
Number of LAEs with only one `real component' & 101 & 4 & 9 & 3 & 117\\
Number of LAEs with two or more `real components' & 100 & 4 & 18 & 9 & 131
\enddata
\end{deluxetable*}

From the above analysis, the fraction of LAEs with two or more components ($408/840=48.6\%$) is very high.  \cite{2016ApJ...819...25K} searched HST/ACS F814W counterparts of 61 ground-based narrowband-selected $z=4.86$ LAEs in COSMOS with a separation of $\leq1\arcsec$, and found that 7 LAEs are not detected in F814W, 8 LAEs have double ACS components, and 46 LAEs have a single component. The fraction of LAEs with two or more components is $8/61=13.1\%$ in their sample, much smaller than our result. We verify our result as follows using JADES data. We download the JADES detection image and the JADES catalog in GOODS-S \citep{2023ApJS..269...16R}. We then match the 509 LAEs in GOODS-S in our sample that are covered by the JADES detection image with the JADES catalog with a separation of $\leq1\arcsec$. The numbers of zero match, one match, and two or more matches are 40, 150, and 319, respectively. If we match the 509 LAEs with our detection catalog, the corresponding numbers are 36, 196, and 277, respectively. Therefore, our result is consistent with the result using the JADES catalog, demonstrating that this high multi-component LAE fraction is reliable. This high fraction, compared with the HST/ACS F814W result in \cite{2016ApJ...819...25K}, could be due to the deeper detection image made by combining deep JWST images in several bands and the wider wavelength range of our detection image from $\sim1.8$ to $\sim5\, \micron$.

In this study, we focus on the 408 LAEs that have two or more counterparts. We visually inspect the images of these LAEs and further refine the sample as follows. In order to obtain robust flux and photo-$z$ measurements for individual components, we require that different components of the same LAEs should be relatively isolated, and each component should be detected at $>5\sigma$ in at least four JWST bands. The final sample consists of 248 LAEs with at least two such components in a circular aperture of 2$\arcsec$ diameter. Figure \ref{fig:LaeEx} shows two LAEs with multiple components in our final sample, including one at $z=5.656$ from \cite{2017ApJ...846..134J} and the other at $z=6.595$ from \cite{2010ApJ...723..869O}. The details of the ground-based images, including the narrowband NB816, $i$ band, and $z$ band images in GOODS-S and the narrowband NB921, $i$ band, and $z$ band images in UDS are described in \cite{2017ApJ...846..134J}. In the ground-based narrowband images, both LAEs seem to have only one component. However, in the JWST images, both LAEs actually show three components. In addition, the rightmost component of the first LAE is clearly a low-redshift foreground object, because it can be seen in the HST F606W band that is at the bluer side of the Ly$\alpha$ emission line at the redshift of the LAE.

\begin{figure}[t]
\plotone{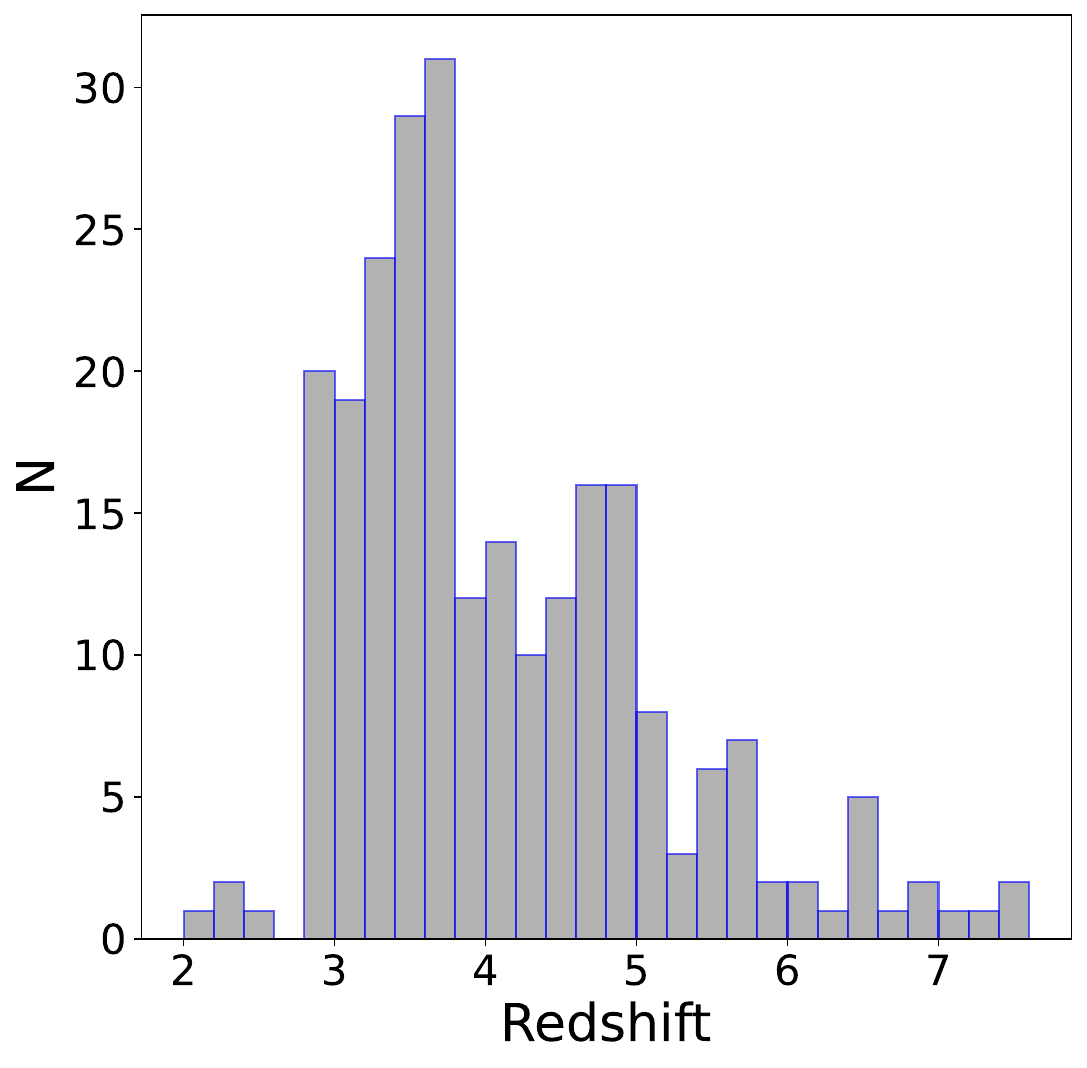}
\caption{Redshift distribution of the LAEs in our final sample.
\label{fig:sampredshift}}
\end{figure}

Table \ref{tab:numLAEComp} shows the numbers of the LAEs and components in the four fields in our final LAE sample. Note that 186 of them ($75\%$; 170 in GOODS-S and 16 in COSMOS) were discovered by the MUSE IFU \citep{2022A&A...659A.183K}. Figure \ref{fig:sampredshift} shows the redshift distribution of these LAEs. Most LAEs are at redshift $\sim3$ to $\sim5$. The catalog of all the LAEs and their components is listed in Table \ref{tab:allinfotable} in the Appendix. The cutout images of all the LAEs in the detection images and the apertures used for all components are shown in Figure \ref{fig:Allcomp1} in the Appendix. 

\subsection{Redshifts of the components in the LAEs} \label{sec:Photoz}

To investigate whether the components of each LAE have the same redshift as the LAE spec-$z$, we use the code {\tt\string EAZY} \citep{2008ApJ...686.1503B} to estimate photo-$z$ for individual components. The template file that we use is {\tt\string eazy\_v1.3.spectra.param} provided by the {\tt\string EAZY} code. We apply the IGM absorption following \cite{2014MNRAS.442.1805I}. Considering possible systematic errors caused by the flux calibration, we add an additional error of $10\%$ of the flux to the photometric errors in quadrature by setting {\tt\string SYS\_ERR} to 0.1. We use a flat prior when estimating photo-$z$. 

\begin{figure}[h]
\epsscale{1.18}
\plotone{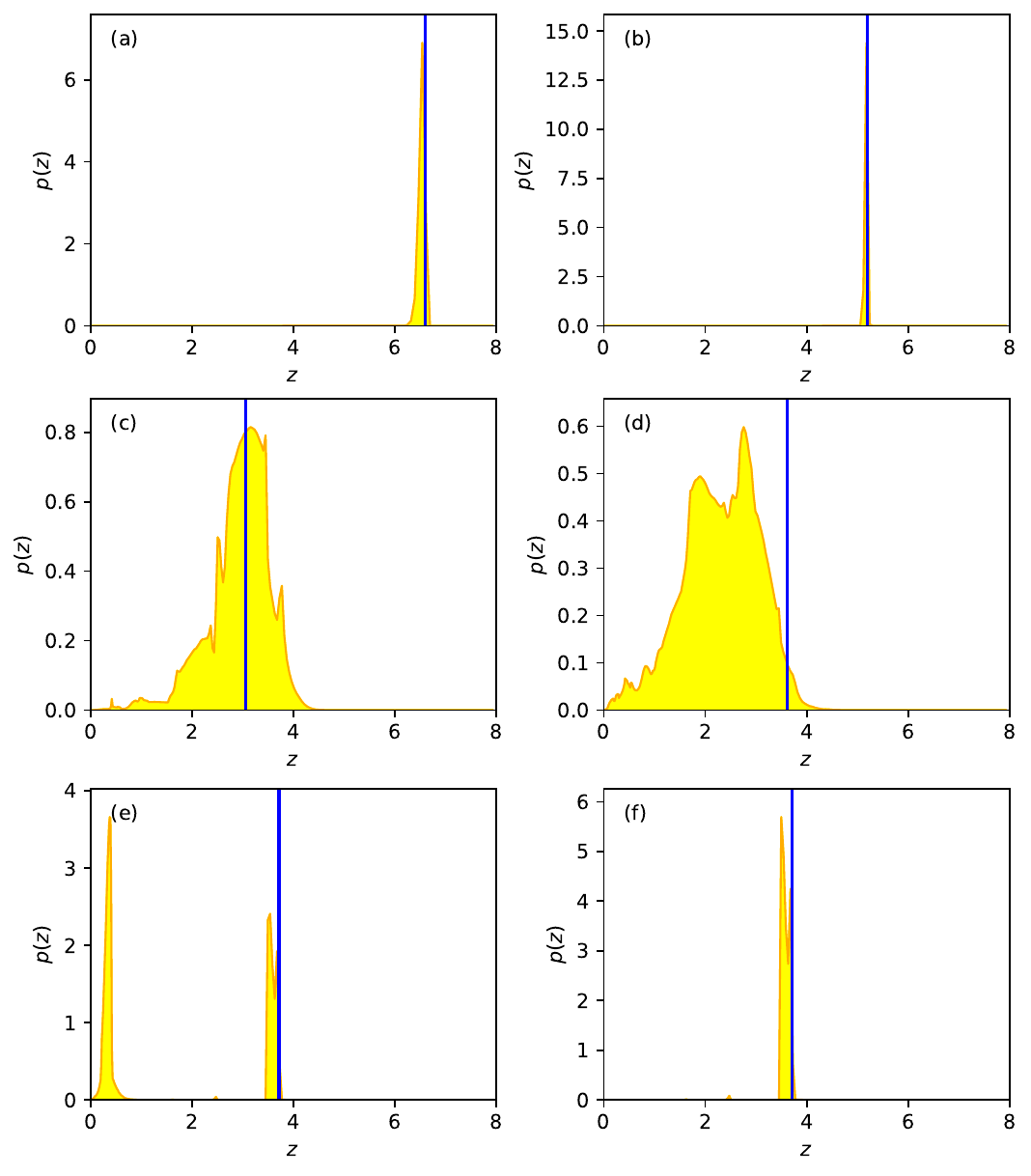}
\caption{Photo-$z$ probability distributions $P(z)$ for some components demonstrating our criteria to choose `real components'. The yellow regions are the $P(z)$ and the blue vertical lines are the spec-$z$ of the LAEs. Panels (a) and (b) show two `real components' with narrow $P(z)$ distributions covering the spec-$z$. Panels (c) and (d) show two components with broad $P(z)$ distributions. The component in panel (c) is regarded as a `real component' because it satisfies {\tt\string l68} $<$ spec-$z$ $<$ {\tt\string u68}, while the component in panel (d) is not. Panels (e) and (f) show the $P(z)$ distributions for the same `real component' with a minimum redshift set to be 0 and 1.5, respectively. 
\label{fig:photoz}}
\end{figure}

After we obtain the posterior photo-$z$ probability distribution $P(z)$ of all components, we compare them with the spec-$z$ of the LAEs and use the following criteria to decide whether the components are consistent with the LAE spec-$z$ (`real components'). These criteria are demonstrated in Figure \ref{fig:photoz}. 
We use the {\tt\string l99}, {\tt\string u99} values given by {\tt\string EAZY} to indicate the $3\sigma$ confidence intervals computed from the probability distributions $P(z)$, and only choose components with {\tt\string l99} $<$ spec-$z$ $<$ {\tt\string u99}. Most of these components show single-peak $P(z)$ distributions with spec-$z$ consistent with the peak (e.g., panels a and b in Figure \ref{fig:photoz}), and these components are considered as `real components'. Some components show very broad $P(z)$ distributions with $1\sigma$ confidence intervals {\tt\string u68} $-$ {\tt\string l68} $>$ 1.0 (e.g., panels c and d in Figure \ref{fig:photoz}). In these cases we only consider those with {\tt\string l68} $<$ spec-$z$ $<$ {\tt\string u68} to be `real components' (e.g., panel c in Figure \ref{fig:photoz}).
Due to the degeneracy between the Lyman break at a high redshift and the Balmer break at a low redshift, the Lyman break at redshift $z$ could be misidentified as the Balmer break at redshift about $(912\sim1216) \times (1+z)/3646$. As a result, some photo-$z$ probability distributions show two peaks, one at a high redshift and the other at a low redshift (e.g., panel e in Figure \ref{fig:photoz}). In these cases, we prefer the high redshift peak if this redshift is consistent with the LAE spec-$z$. For these components, we rerun {\tt\string EAZY} with the minimum redshift of 1.5 (e.g., panel f in Figure \ref{fig:photoz}). 

\begin{figure}[h]
\epsscale{1.0}
\plotone{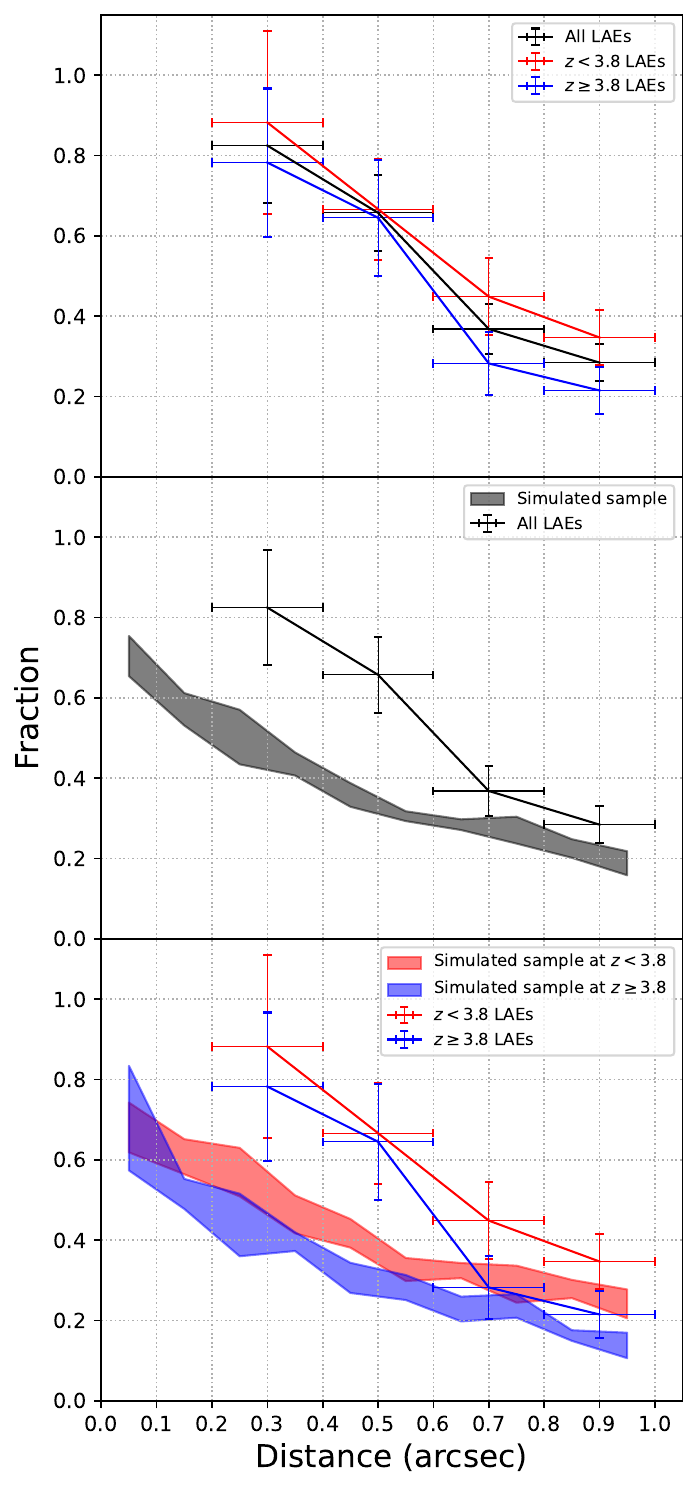}
\caption{Fraction of `real component' as a function of projected distance to the central `real component' for the LAE sample and simulated sample. The top panel shows the fractions for the entire LAE sample (black), the $z<3.8$ LAE subsample, and the $z\geqslant3.8$ LAE subsample. The error bar in the y-direction denotes the $1\sigma$ Poisson error. The error bar in the x-direction denotes the range of distance in this bin. The gray shaded area in the middle panel shows the range between the 25th and 75th percentiles of the fractions for the entire simulated sample in the 8 simulated catalogs. The red/blue shaded area in the bottom panel shows the range between the 25th and 75th percentiles of the fractions for the simulated sample at $z<3.8$ and $z\geqslant3.8$ in the 8 simulated catalogs.
\label{fig:realcompfrac}}
\end{figure}

These criteria classify each component as either a `real component' at the LAE redshift, or a foreground/background object aligning coincidentally with the line of sight of the LAE. We find that 403/593=68.0\% of all the components are `real components', and each LAE has at least one `real component'. We also estimate the fraction of `real component' as a function of projected distance to the central `real component'. Here a central `real component' is defined as a `real component' that is the closest to the LAE coordinate. We then calculate the projected distances of all other components to it. We do not have any LAEs with projected distances less than $0\farcs2$ in our sample, because the projected distance between two components is too small so that they cannot be deblended in the detection image, and thus such LAEs were excluded during the construction of the sample. At projected distances larger than $0\farcs2$, the fraction of `real component' basically decreases with increasing projected distance, from $\sim80\%$ at $0\farcs2-0\farcs4$ to $\sim30\%$ at $0\farcs8-1\farcs0$. This is reasonable because larger projected distances mean larger areas and greater chances to intersect with foreground/background objects. Also note that this fraction is $100\%$ at $0\farcs2-0\farcs3$. Table \ref{tab:frac_dist} and Figure \ref{fig:realcompfrac} show the result of the above calculations. The $1\sigma$ Poisson errors of the fractions are shown in Figure \ref{fig:realcompfrac}.

In order to check the redshift dependence of the real-component fractions, we divide the LAE sample into two subsamples with similar numbers of LAEs, including one at low redshift ($z<3.8$, 127 LAEs) and the other at high redshift ($z\geqslant3.8$, 121 LAEs). Table \ref{tab:frac_dist} and Figure \ref{fig:realcompfrac} also show the results in the two subsamples. We find that both subsamples show similar trends compared with the whole LAE sample. At small projected distances ($<0\farcs6$), the fractions in the two subsamples are comparable (around $\sim65\%$ to $\sim90\%$). At larger projected distances ($\gtrsim0\farcs6$), the $z<3.8$ subsample has higher fractions ($\sim10\%$ to $\sim20\%$ higher) than the $z\geqslant3.8$ subsample with discrepancies larger than $1\sigma$ Poisson error. The reason can be twofold. On one hand, low-redshift galaxies have relatively more higher-redshift interlopers, whereas high-redshift galaxies have relatively more lower-redshift interlopers. Under the same detection limit, interlopers at lower redshift are more likely to be detected, causing a lower fraction of the $z\geqslant3.8$ subsample at all projected distances. On the other hand, galaxies at higher redshift have stronger clustering \citep[e.g., Figure 9 and Figure 10 in][]{2022MNRAS.515.5416Y}, which could increase the fractions of the $z\geqslant3.8$ subsample.

\begin{deluxetable*}{cccc|cccc}[t]
\tabletypesize{\footnotesize}
\tablewidth{0pt}
\tablecaption{Fraction of `real component' as a function of projected distance to the central `real component'. \label{tab:frac_dist}}
\tablehead{
\colhead{Distance} & \multicolumn{3}{c}{LAEs} & \colhead{Distance} & \multicolumn{3}{c}{Simulated sample} \\
\cline{2-4}
\cline{6-8}
\colhead{(arcsec)} & \colhead{All} & \colhead{$z<3.8$} & \colhead{$z\geqslant3.8$} & \colhead{(arcsec)} & \colhead{All} & \colhead{$z<3.8$} & \colhead{$z\geqslant3.8$}
}
\startdata
$0 - 0.2$   & ...           & ...          & ...          & $0 - 0.1$ & 71\% & 68\% & 69\% \\
            &               &              &              & $0.1 - 0.2$ & 57\% & 62\% & 51\% \\
$0.2 - 0.4$ & 33/40 = 83\%  & 15/17 = 88\% & 18/23 = 78\% & $0.2 - 0.3$ & 54\% & 61\% & 45\% \\
            &               &              &              & $0.3 - 0.4$ & 43\% & 46\% & 41\% \\
$0.4 - 0.6$ & 48/73 = 66\%  & 28/42 = 67\% & 20/31 = 65\% & $0.4 - 0.5$ & 35\% & 39\% & 31\% \\
            &               &              &              & $0.5 - 0.6$ & 30\% & 33\% & 28\% \\
$0.6 - 0.8$ & 35/95 = 37\%  & 22/49 = 45\% & 13/46 = 28\% & $0.6 - 0.7$ & 28\% & 33\% & 22\% \\
            &               &              &              & $0.7 - 0.8$ & 28\% & 30\% & 24\% \\
$0.8 - 1.0$ & 39/137 = 28\% & 25/72 = 35\% & 14/65 = 22\% & $0.8 - 0.9$ & 23\% & 27\% & 16\% \\
            &               &              &              & $0.9 - 1.0$ & 19\% & 24\% & 14\% 
\enddata
\tablecomments{`...' means there is no such component. The fractions for the simulated sample are the medians of the 8 simulated catalogs.}
\end{deluxetable*}

To further explain these trends, we compare the observational results with the theoretical predictions made with the Santa Cruz semi-analytic model (SAM) for galaxy formation \citep{2022MNRAS.515.5416Y}. The SAM has been shown to reproduce a variety of observational constraints at high redshift, including UV LFs, stellar mass functions, star formation rate (SFR) functions, stellar-to-halo mass ratios, and two-point correlation functions (2PCFs). Since most LAEs in our sample are in the GOODS-S field, we use the 8 wide-field lightcones overlapping the GOODS-S field produced by the SAM model. These wide-field lightcones provide simulated magnitudes in most JWST NIRCam bands, and resolve galaxies down to stellar mass $\sim10^7 \, M_{\odot}$, which is similar to the smallest stellar masses in our galaxies (see Section \ref{sec:SED}). 
The observational results depend on the object selection criteria and survey depth. Because the 8 simulated catalogs do not provide Ly$\alpha$ flux or EW, here we select galaxies with specific star formation rate (sSFR) $>10^{-8.1} \, \rm yr^{-1}$ which is satisfied by all Ly$\alpha$ emitting components in Section \ref{sec:SED}. We randomly select 1000 galaxies in each simulated catalog with sSFR $>10^{-8.1} \, \rm yr^{-1}$ and with two or more components in a circular aperture of $2\arcsec$ diameter. We further require that the 1000 selected galaxies have a redshift distribution similar to that of the LAE sample and that every component has magnitudes in at least four JWST bands brighter than the $5\sigma$ detection limits of our images. Considering the uncertainty of the photo-$z$, we regard components with redshift deviations from the selected central components smaller than 0.1 as `real components'. Then we calculate the fractions of `real components' as a function of projected distances to the central components. We also split the 1000 selected galaxies in each catalog into $z<3.8$ and $z\geqslant3.8$ subsamples and calculate the fractions in each subsample.

Table \ref{tab:frac_dist} and Figure \ref{fig:realcompfrac} list the median values, the 25th percentile, and the 75th percentile of the fractions in the 8 simulated catalogs. The theoretical predictions of the fractions for the entire sample are apparently lower than those of the LAE sample at projected distances $<0\farcs6$, which is also true for the two subsamples. At projected distances from $0\farcs6$ to $1\farcs0$, the theoretical predictions are still lower than the observations, but the discrepancies are smaller than those at projected distances $<0\farcs6$. 
There are several possible reasons for the discrepancies. First, the virial radii of the host halos of the selected simulated galaxies are all above 10 kpc, which corresponds to more than $1\arcsec$ projected distance at redshift $z\sim2$ to $z\sim7$. This means that the clustering on small scales is dominated by galaxies that reside within the same host halo.  \cite{2022MNRAS.515.5416Y} noted that the properties and positions of satellite galaxies are particularly uncertain in the SAM model. As a result, the predicted fractions can be uncertain on small scales. Indeed, \cite{2022MNRAS.515.5416Y} showed that at $1.25\lesssim z \lesssim4.5$, the predicted 2PCFs are different from the CANDELS observational results on small scales. Second, as mentioned in \cite{2022MNRAS.515.5416Y}, photometry in the simulated catalogs includes only contributions from the stellar continuum and ignores the emission from nebular lines and continuum. LAEs are typically strong emission line galaxies. Therefore, ignoring nebular emission may affect the sample selection and effective survey depth. For example, some faint and young companion star-forming galaxies of the central components would be brighter in some JWST bands if line emission is included, so they would have been selected and thus increase the fractions. Third, the theoretical predictions do not consider the situations in which a foreground galaxy blocks and/or blends with a background galaxy so that they cannot be separated in the images. When we constructed our LAE sample, we excluded some LAEs that are severely affected by bright nearby objects. This increases fractions at smaller projected distances in the LAE sample, especially taking into account the typically small sizes of LAEs \citep{2016ApJ...819...25K}. Finally, our sample selection with sSFR $>10^{-8.1} \, \rm yr^{-1}$ will also include LBGs without Ly$\alpha$ emission, whose clustering properties could be different from LAEs with Ly$\alpha$ luminosities $\gtrsim 10^{42.2} \,\rm erg\, s^{-1}$ in our sample.

\subsection{Properties of the LAE components} \label{sec:SED}

\begin{deluxetable*}{ccc}[t]
\tabletypesize{\scriptsize}
\tablewidth{0pt}
\tablecaption{{\tt\string CIGALE} configurations. \label{tab:SEDConfig}}
\tablehead{
\colhead{Parameter} & \colhead{Symbol} & \colhead{Range} 
}
\startdata
 & \textit{Delayed star formation history} & \\
\textit{e}-folding time of main stellar population (Myr) & $\tau_{\mathrm{main}}$ & 10, 25, 50, 100, 250, 500, 1000, 2500, 5000, 10000, 100000\\
Age of main stellar population (Myr) & age$_{\mathrm{main}}$ & 20, 50, 80, 100, 250, 500, 1000, 2500\\
\hline
 & \textit{Stellar populations} & \\
 \multicolumn{3}{c}{\centering{Stellar population synthesis models from \cite{2003MNRAS.344.1000B}}} \\
\hline
Initial mass function & IMF & Chabrier\\
Metallicity & Z$_{\mathrm{star}}$ & 0.0001, 0.0004, 0.004, 0.008, 0.02\\
\hline
 & \textit{Nebular emission} & \\
Ionization parameter & logU & $-$3.5, $-$3.0, $-$2.5, $-$2.0, $-$1.5, $-$1.0\\
Gas metallicity & Z$_{\mathrm{gas}}$ & 0.0004, 0.001, 0.005, 0.009, 0.014, 0.02, 0.025, 0.03, 0.041, 0.051\\
Electron density & n$_e$ & 10, 100, 1000\\
\hline
 & \textit{Modified starburst dust attenuation law} &\\
Color excess of nebular lines & E(B-V)$_{\mathrm{lines}}$ & 0.05, 0.1, 0.2, 0.3, 0.4, 0.5\\
Reduction factor to compute E(B-V)$_{\mathrm{stellar}}$ & E(B-V)$_{\mathrm{factor}}$ & 0.3, 0.6, 0.9\\
Amplitude of the 217.5 nm UV bump & uv\_bump\_amplitude & 0, 1.5, 3.0\\
Power-law slope & $\delta$ & $-0.5$, $-0.4$, $-0.3$, $-0.2$, $-0.1$, 0
\enddata
\end{deluxetable*}

With photo-$z$ we have determined whether a component is a `real component' or a foreground/background object. We divide our LAE sample into two subsamples, Subsample 1 and Subsample 2. Subsample 1 contains 117 LAEs with only one `real component' and Subsample 2 contains 131 LAEs with two or more `real components'. The numbers of LAEs in two Subsamples are listed in Table \ref{tab:numLAEComp}. The Ly$\alpha$ photons of the LAEs in Subsample 1 must be emitted by the only one `real component'. For LAEs in Subsample 2, we do not know which component emits Ly$\alpha$ photons, even with help from ground-based narrowband images (like the two cases shown in Figure \ref{fig:LaeEx}). Since the Ly$\alpha$ line is a resonant line, the radiative transfer of Ly$\alpha$ photons is complex, depending on the constitution, geometry, and kinematics of the interstellar medium (ISM). In principle, deep and high-spatial-resolution spectra covering Ly$\alpha$ can determine the origin of the Ly$\alpha$ line emission in Subsample 2. Here we  measure the physical properties of the individual LAE components with SED modeling, and estimate which components likely produce Ly$\alpha$ line emission. We use {\tt\string CIGALE} \citep{2019A&A...622A.103B} to perform the SED fitting with configuration values listed in Table \ref{tab:SEDConfig}. Considering the complex Ly$\alpha$ radiative transfer and the stochastic IGM absorption, we only use bands redward of Ly$\alpha$ to fit the SED. We turn on nebular emission since most LAEs have very strong nebular emission lines that can even largely affect broadband magnitudes \citep[e.g.,][]{2024arXiv240611997K}. One of the advantages of our LAE sample is that we can fix the redshift to the spec-$z$ for the SED modeling, which reduces the uncertainties of the derived physical properties. 

\begin{figure}[t]
\epsscale{1.18}
\plotone{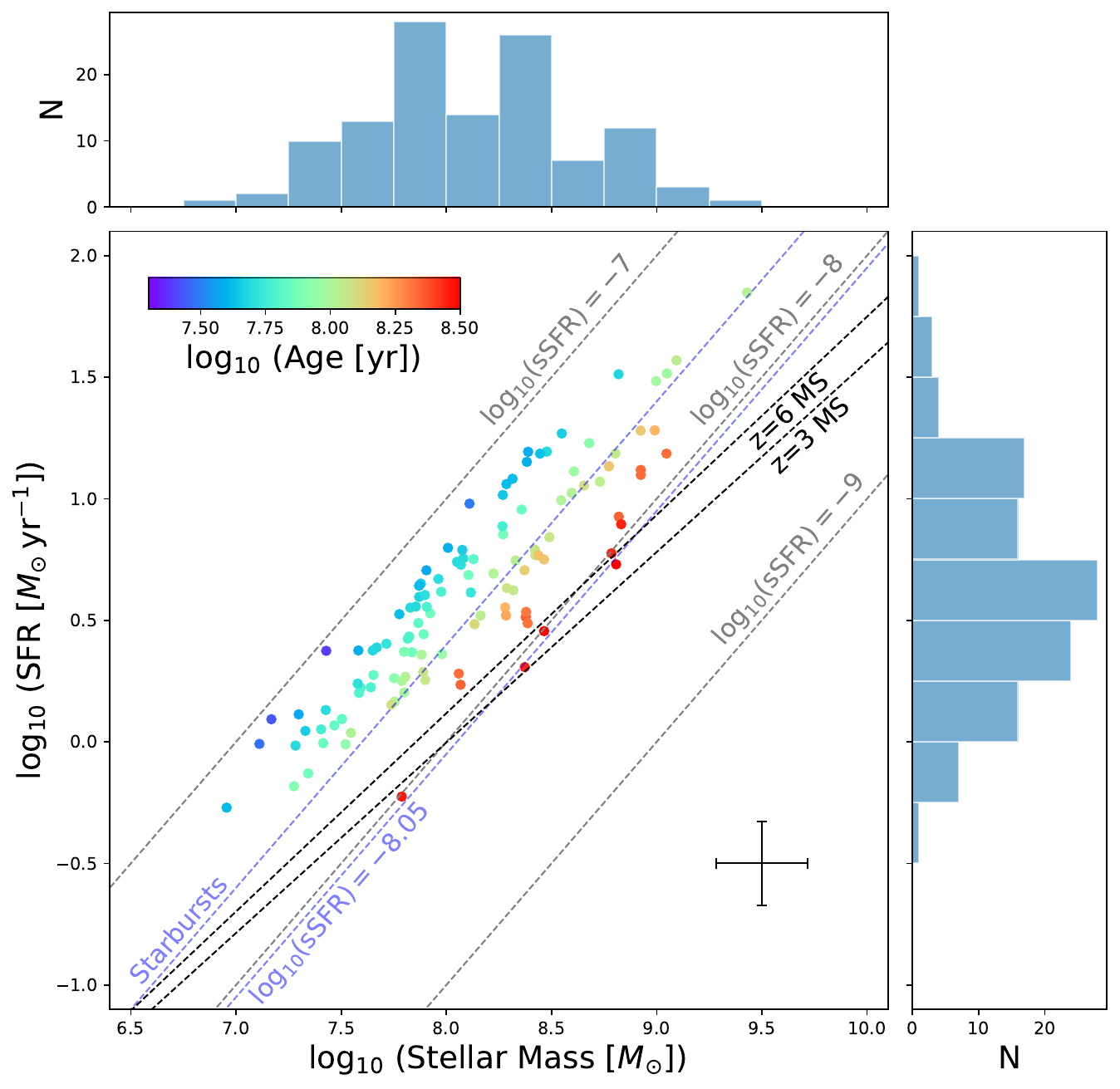}
\caption{SFR - $M_*$ diagram for `real components' in Subsample 1, color-coded by age. The upper and right panels show the histograms of $M_*$ and SFR, respectively. The three grey dashed lines refer to log$_{10}\,$(sSFR)$\,=-7$, $-8$ and $-9$, respectively. The two black dashed lines indicate the main sequences of galaxies at $z=3$ and $z=6$ from \cite{2014ApJS..214...15S}, respectively. The upper blue dashed line (log$_{10}\,$(sSFR)$\,=-7.6$) is the separation between starburst and SF valley defined in \cite{2017ApJ...849...45C}. The lower blue dashed line (log$_{10}\,$(sSFR)$\,=-8.05$) is the separation between SF valley and main sequence defined in \cite{2017ApJ...849...45C}. The median errors of SFR and $M_*$ are shown in the lower right corner. 
\label{fig:sfr_mass_only_one}}
\end{figure}

\begin{figure}[t]
\epsscale{0.9}
\plotone{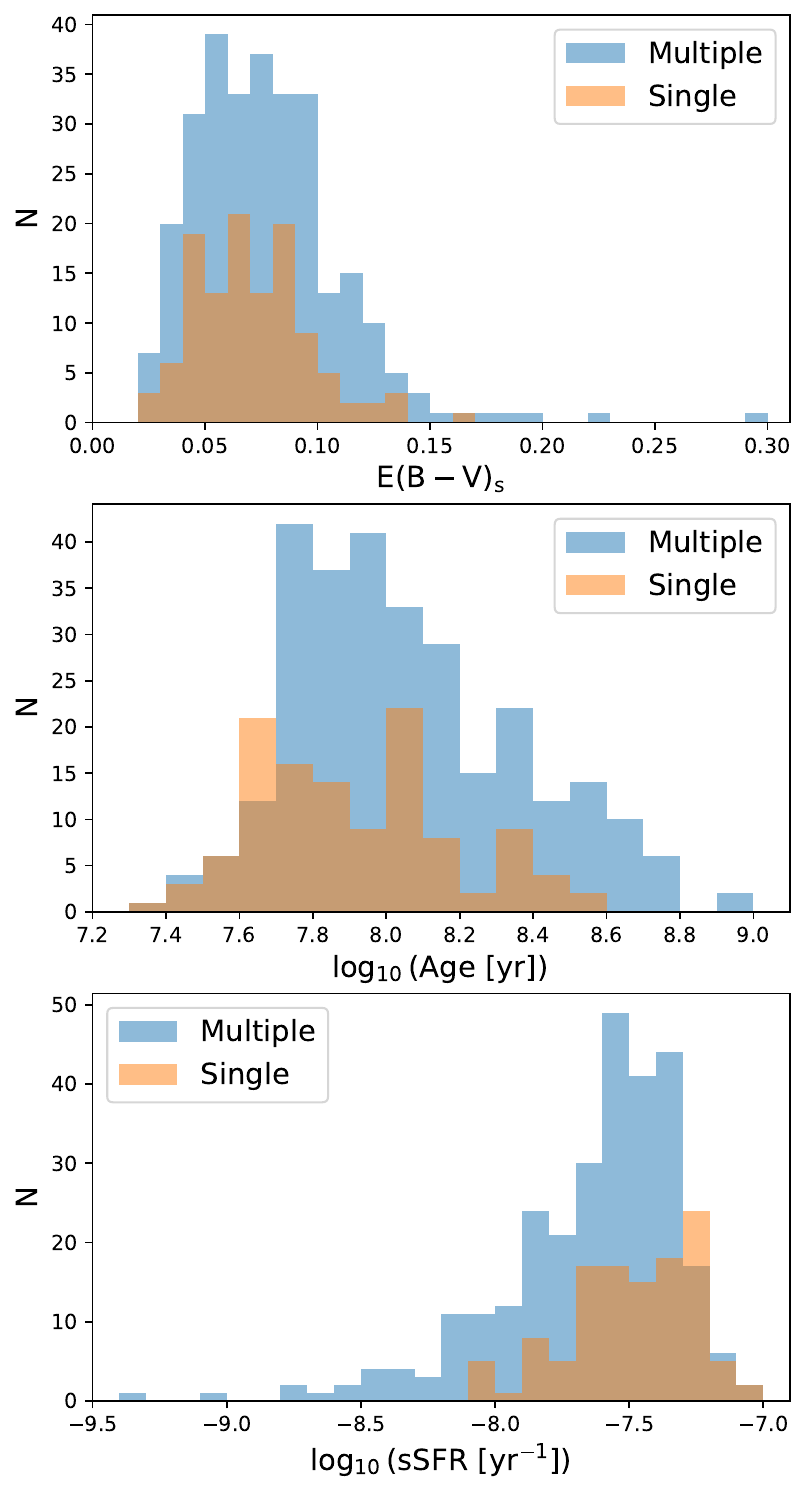}
\caption{Distributions of $\rm E(B-V)_s$, age and sSFR for the `real components' in Subsample 2 (blue) and Subsample 1 (orange).
\label{fig:ebvsagessfr}}
\end{figure}

Figure \ref{fig:sfr_mass_only_one} shows the SFR - stellar mass ($M_{*}$) diagram, color-coded by age, and the distributions of SFR and $M_{*}$ for the Ly$\alpha$ emitting components in Subsample 1. Figure \ref{fig:ebvsagessfr} shows the distributions of $\rm E(B-V)_s$, age, and sSFR for these components. These LAEs show a median $M_{*}$ of $10^{8.07} \, M_{\odot}$, a median SFR of $3.95 \, M_{\odot} \, \rm yr^{-1}$, a median age of $10^{7.87} \, $yr, a median $\rm E(B-V)_s$ of 0.068, and a median sSFR of $10^{-7.45} \, \rm yr^{-1}$, consistent with the overall picture that LAEs are young, low-mass, low dust-extinction, star-forming galaxies. We also plot the main sequence (MS) of star-forming galaxies at $z=3$ and $z=6$ from \cite{2014ApJS..214...15S}. All but one of these $z\sim2$ to $z\sim7$ LAEs lie above the MS of galaxies, and all of them have sSFR $>10^{-8.1} \, \rm yr^{-1}$, meaning that these LAEs have stronger star-forming activities than normal MS galaxies. \cite{2017ApJ...849...45C} studied H$\alpha$ emitters (HAEs) at redshift $3.9<z<4.9$ and found that these HAEs can be divided into three categories based on their sSFR: Starburst (sSFR $>10^{-7.6} \, \rm yr^{-1}$), SF valley ($10^{-7.6} \, \rm yr^{-1}$ $>$ sSFR $>10^{-8.05} \, \rm yr^{-1}$), and MS (sSFR $<10^{-8.05} \, \rm yr^{-1}$). The lowest sSFR of the LAEs in Subsample 1 (sSFR $=10^{-8.08} \, \rm yr^{-1}$) is consistent with the separation between the SF valley and the MS of HAEs in \cite{2017ApJ...849...45C}, suggesting that most of these LAEs are starburst galaxies. Young OB stars in these LAEs emit a large amount of ionizing photons, which partly turn into Ly$\alpha$ emission later and result in the large Ly$\alpha$ EW of LAEs. \cite{2024ApJ...963...97I} also found that $75\%$ of their LAEs populate the starburst region in the SFR - $M_{*}$ diagram. The remaining $25\%$ of their LAEs are older, more massive, and consistent with the MS of star-forming galaxies. The reason why almost all of our LAEs are in the starburst region is that our sample selection criteria selected LAEs with stronger Ly$\alpha$ luminosity ($\gtrsim 10^{42.2} \,\rm erg\, s^{-1}$). The older LAEs in \cite{2024ApJ...963...97I} have a median Ly$\alpha$ luminosity of $10^{41.87} \,\rm erg\, s^{-1}$, thus have been neglected.

\begin{figure}[t]
\epsscale{1.18}
\plotone{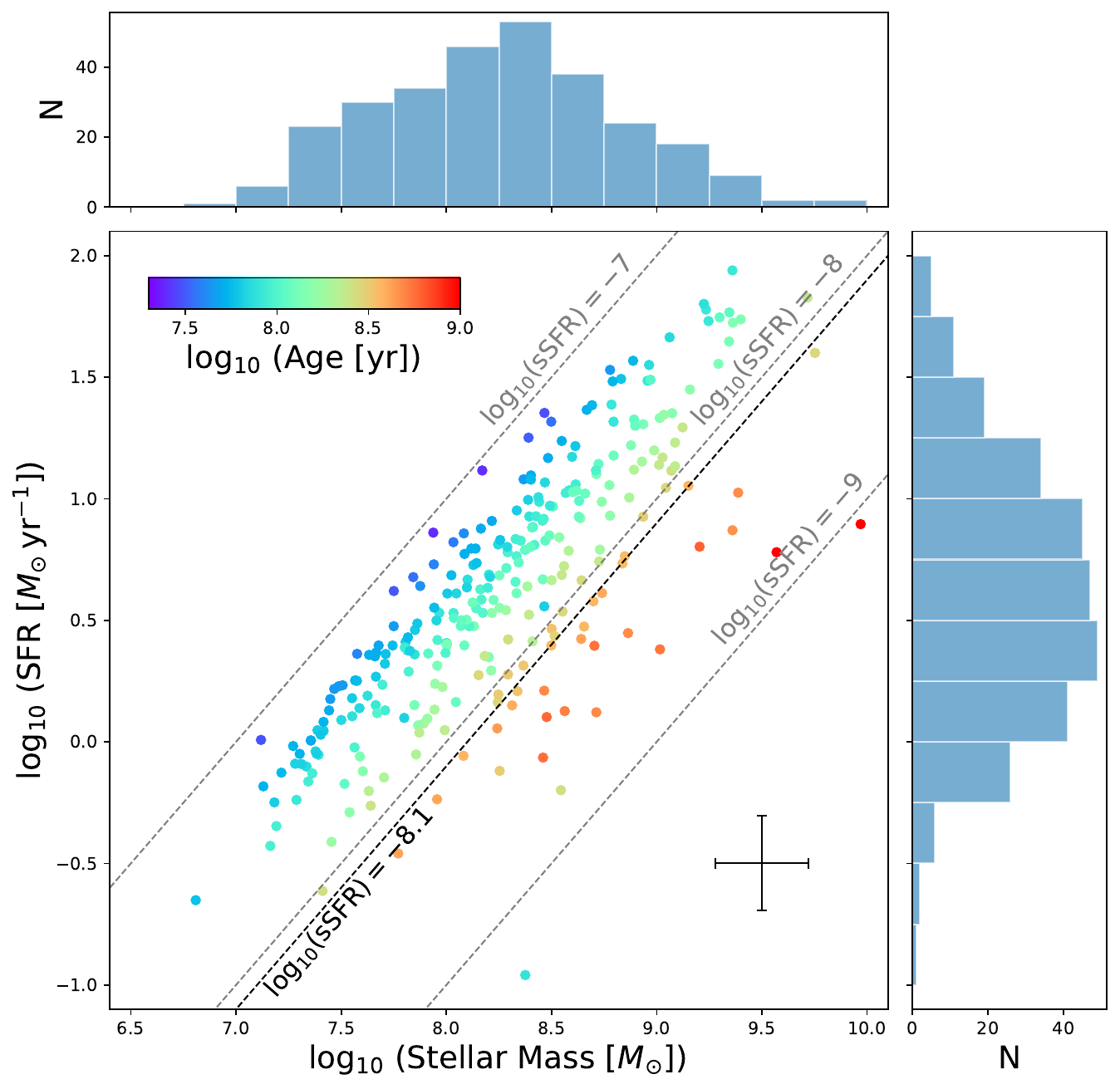}
\caption{SFR - $M_*$ diagram for `real components' in Subsample 2, color-coded by age. The upper and right panels show the histograms of $M_*$ and SFR, respectively. The three grey dashed lines refer to log$_{10}\,$(sSFR)$\,=-7$, $-8$ and $-9$, respectively. The components above the black dashed line (log$_{10}\,$(sSFR)$\,=-8.1$) likely emit Ly$\alpha$ photons. Median errors of SFR and $M_*$ are shown in the lower right corner. 
\label{fig:sfr_mass_more_comp}}
\end{figure}

\begin{figure*}[t]
\epsscale{1.18}
\plotone{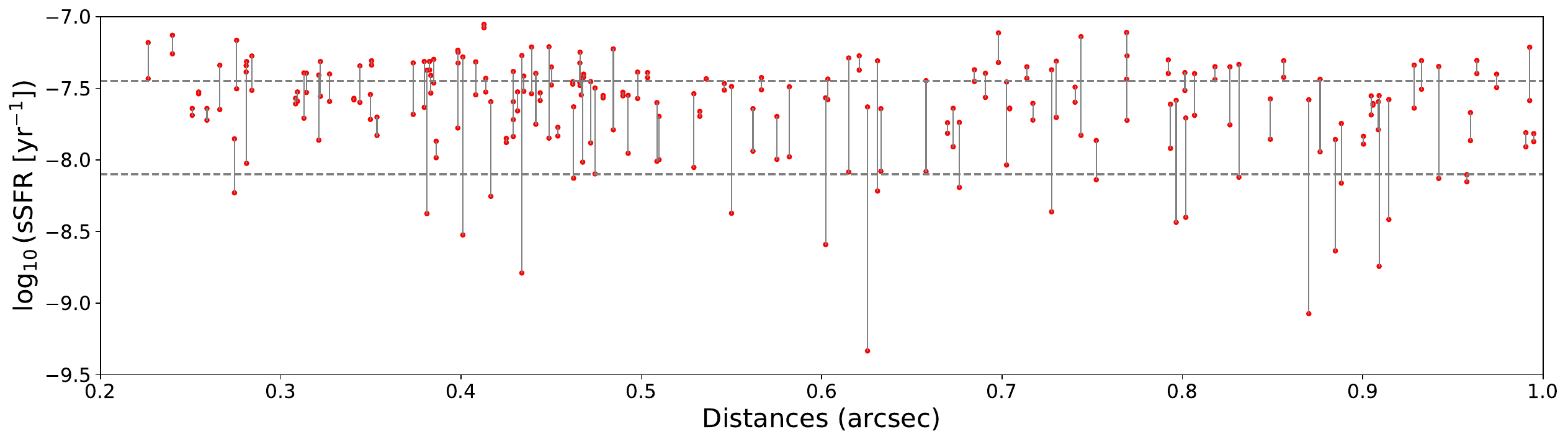}
\caption{The sSFR values for two `real components' in each LAE in Subsample 2 as a function of the projected distance between the two `real components'. Each vertical line connects two components in the same LAE. For LAEs with more than two `real components', we only plot the central one and the one nearest to it. The upper gray dashed line is the median sSFR for the Ly$\alpha$ emitting components in Subsample 1 and the lower gray dashed line is sSFR $=10^{-8.1} \, \rm yr^{-1}$. 
\label{fig:comparecomp}}
\end{figure*}

Figure \ref{fig:sfr_mass_more_comp} shows the SFR - $M_{*}$ diagram, color-coded by age, and the distributions of SFR and $M_{*}$ for the `real components' in Subsample 2. Figure \ref{fig:ebvsagessfr} shows the distributions of $\rm E(B-V)_s$, age, and sSFR for these components. If we apply the criterion sSFR $>10^{-8.1} \, \rm yr^{-1}$ satisfied by all Ly$\alpha$ emitting components in Subsample 1 to the 286 `real components' in Subsample 2, we find that 257 of them (i.e., $257/286=90\%$) meet this criterion. Figure \ref{fig:comparecomp} shows the sSFR for two `real components' in each LAE in Subsample 2 as a function of the projected distance between the two `real components'. For LAEs with more than two `real components', we only plot the central one and the one nearest to it. We can see that for each LAE, there is at least one `real component' with sSFR $>10^{-8.1} \, \rm yr^{-1}$. Therefore, $90\%$ of the `real components' in Subsample 2 have stronger star-forming activities than normal MS galaxies, just like the Ly$\alpha$ emitting components in Subsample 1. \cite{2024ApJ...963...97I} found that there is no statistical difference in sSFR between LAEs and a mass-matched sample of LBGs, so the components with sSFR $>10^{-8.1} \, \rm yr^{-1}$ can also be LBGs that do not have Ly$\alpha$ emission. Finally, we do not see any correlation between the sSFR and the projected distance from $0\farcs2$ to $1\farcs0$ in Figure \ref{fig:comparecomp}. The sSFR of these sSFR $>10^{-8.1} \, \rm yr^{-1}$ `real components' in Subsample 2 are consistent with the median of the sSFR of the Ly$\alpha$ emitting components in Subsample 1. This likely means that galaxy interaction is not the main driver of the strong star-forming activities in LAEs.

\section{Discussion} \label{sec:discussion}

In the previous sections, we have shown that many LAEs have multiple components in a circular aperture of $2\arcsec$ in diameter. Some of these components are at the LAE redshifts, and the others are foreground/background interlopers. In this section, we discuss how this affects the results of ground-based LAE studies. Ground-based LAE studies usually use a circular aperture of $2\arcsec$ to do photometry, so the measured flux could be severely affected. Since our LAE sample is constructed by compiling spec-confirmed LAEs in the literature, it is not a complete or uniform sample. Therefore, we are not able to discuss the effect in a quantitative way. Instead, we qualitatively show some overall trends based on our multi-component LAE sample. 

To mimic the atmospheric seeing effect on ground-based observations, we calculate a homogenization kernel between the PSF in each filter constructed in Section \ref{sec:photometry} and a Gaussian PSF with FWHM of $0\farcs8$, and then convolve these kernels with the space-based images to get simulated ground-based images. For each LAE in the sample, we use {\tt\string SExtractor} to measure its simulated ground-based flux in all available HST and JWST bands with the same detection image used in Section \ref{sec:photometry} and with a circular aperture of $2\arcsec$ centered on the central `real component' defined in Section \ref{sec:Photoz}. By setting {\tt\string MASK\_TYPE} = NONE in the configuration file, we obtain the flux that would be measured in a ground-based LAE survey. We do aperture correction on the measured flux by assuming a point-source model, i.e., by multiplying the flux with the ratio between the total flux of a Gaussian PSF with FWHM of $0\farcs8$ and the flux that is within a circular aperture of $2\arcsec$ centered on the Gaussian PSF. We use HST F435W, HST F606W, HST F814W, JWST F090W, and JWST F115W magnitudes as the UV magnitudes for LAEs at $2<z<2.2$, $2.2<z<3.0$, $3.0<z<4.6$, $4.6<z<5.65$, and $5.65<z<7.6$, respectively. We also use {\tt\string CIGALE} and the new flux to do SED fitting with the same configurations as in Section 3.3 and obtain the physical properties of the LAEs. In the following discussion, we use `$2\arcsec$' to denote the new measurements. 

\begin{figure*}[t]
\epsscale{1.18}
\plotone{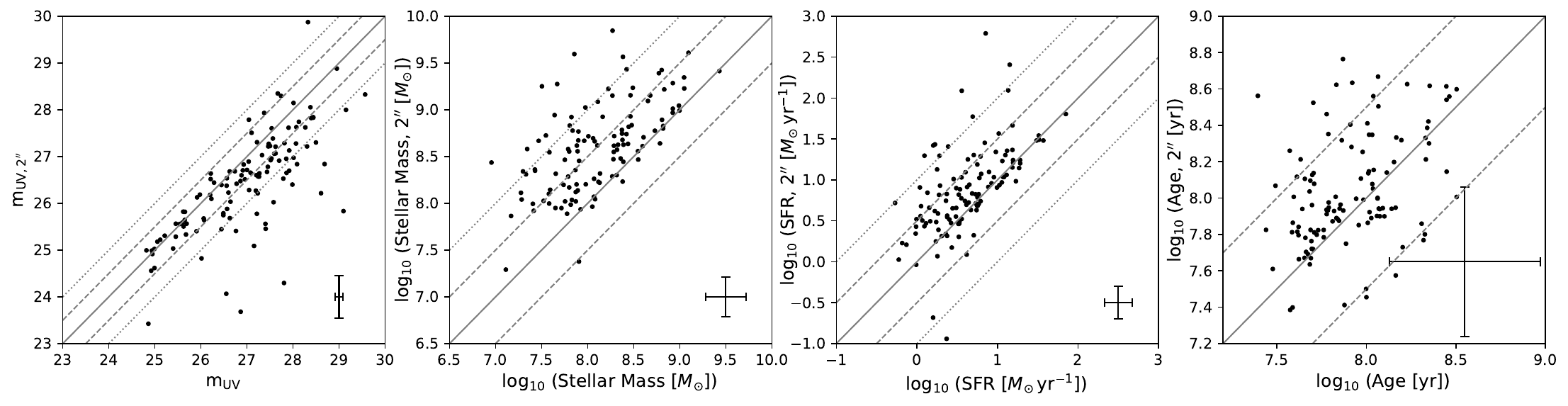}
\caption{Comparisons of UV magnitudes, stellar mass, SFR, and age for LAEs in Subsample 1 between the space-based measurements and the simulated ground-based measurements. Y axis refers to a circular aperture of $2\arcsec$ diameter on the simulated ground-based observations and X axis refers to the space-based measurements in previous sections. Gray solid lines are one-to-one lines, gray dashed lines denote a difference of 0.5 mag or 0.5 dex, and gray dotted lines denote a difference of 1.0 mag or 1.0 dex. Median errors are shown in the lower right corner of all panels. $\rm m_{UV,2\arcsec}$ has larger magnitude errors than $\rm m_{UV}$ because of using a larger aperture. Age measurements have very large errors and are not so reliable.
\label{fig:circsingle}}
\end{figure*}

We first focus on the 117 LAEs in Subsample 1. Since the central `real component' is the only component emitting Ly$\alpha$ photons in this subsample, the flux from other components in the circular aperture of $2\arcsec$ is contamination in ground-based LAE studies. Figure \ref{fig:circsingle} compares the UV magnitude, stellar mass, SFR, and age of the new measurements with the previous measurements. As expected, the new magnitudes $\rm m_{UV,2\arcsec}$ are typically brighter than $\rm m_{UV}$: 53 LAEs have $\Delta \rm m_{UV}=m_{UV}-m_{UV,2\arcsec} \geqslant 0.5$ mag and 25 LAEs have $\Delta \rm m_{UV}\geqslant 1.0$ mag. Consequently, the new stellar masses $\rm M_{*,2\arcsec}$ are mostly larger than $\rm M_*$: 51 LAEs have $\Delta \rm log_{10} M_*=log_{10} M_{*,2\arcsec}-log_{10} M_* \geqslant 0.5$, and 15 LAEs have $\Delta \rm log_{10} M_* \geqslant 1.0$. Similarly, the new SFRs $\rm SFR_{2\arcsec}$ are larger than SFR: 25 LAEs have $\Delta \rm log_{10} SFR=log_{10} SFR_{2\arcsec}-log_{10} SFR \geqslant 0.5$ and 9 LAEs have $\Delta \rm log_{10} SFR \geqslant 1.0$. The age measurements have larger uncertainties, and we cannot draw any conclusions.

\begin{figure*}[t]
\epsscale{1.18}
\plotone{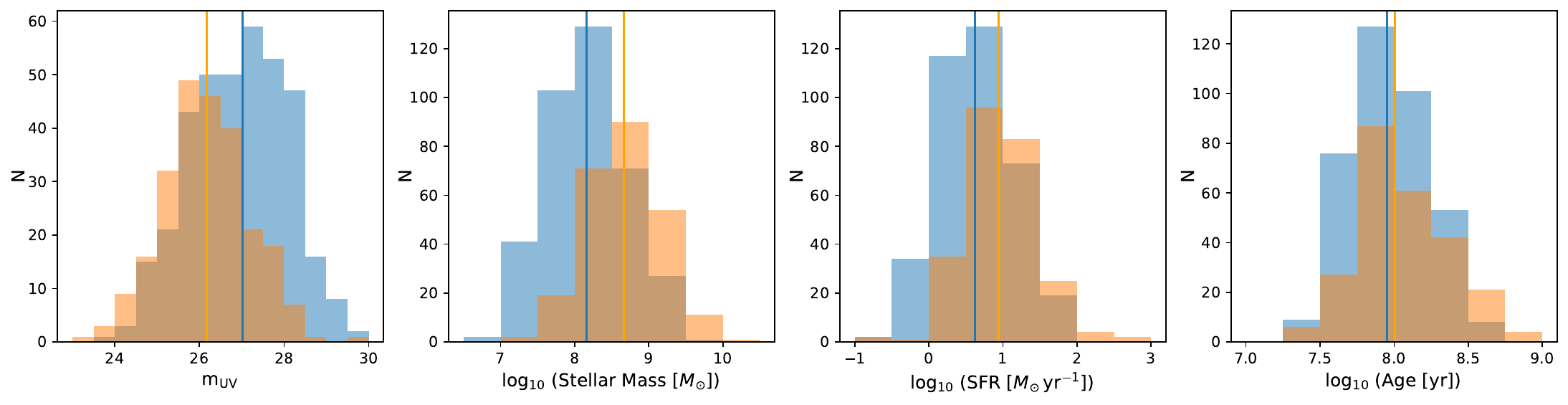}
\caption{Distributions of UV magnitudes, stellar mass, SFR, and age for all multi-component LAEs in our sample in the space-based measurements and the simulated ground-based measurements. Orange histograms refer to a circular aperture of $2\arcsec$ diameter on the simulated ground-based observations and blue histograms refer to the space-based measurements in previous sections. We assume all components with sSFR$>10^{-8.1} \, \rm yr^{-1}$ are emitting Ly$\alpha$ photons for LAEs in the `multiple subsample'. Orange and blue vertical lines denote median values. 
\label{fig:circall}}
\end{figure*}

For LAEs in Subsample 2, we cannot determine which components are emitting Ly$\alpha$ photons, so we assume that all components with sSFR$>10^{-8.1} \, \rm yr^{-1}$ are Ly$\alpha$ emitting components in the following discussion. Using a larger aperture not only includes contamination from foreground/background interlopers in the flux measurement, but also combines the flux from two or more Ly$\alpha$ emitting components in LAE samples. These potentially affect the measurement of galaxy LF, stellar mass, and other properties of LAE samples such as photometric redshifts. Figure \ref{fig:circall} compares the distributions of the newly measured UV magnitudes, stellar mass, SFR, and age for all LAEs in Subsamples 1 and 2 with the  measurements in Section 3. We find that using a circular aperture of $2\arcsec$ results in a $\sim0.85$ mag brighter median UV magnitude, $\sim0.5$ dex larger median stellar mass, and $\sim0.3$ dex larger median SFR in our LAE sample. Given that the fraction of multi-component LAEs in our initial LAE sample is nearly 50\%, these effects would largely change some of the previous results based on previous ground-based studies. More details can be obtained from a well-define LAE sample in the future.

\section{Summary} \label{sec:summary}

We have investigated LAEs with multiple components in GOODS-S, GOODS-N, COSMOS, and UDS using HST and JWST images. We used HST HLF images in GOODS-S/GOODS-N and HST CANDELS images in COSMOS/UDS. We reduced JWST NIRCam images, combined them, and made mosaics in the four fields. By compiling spectroscopically confirmed LAEs from $z\sim2$ to $z\sim7$ in the literature and matching these LAEs with our detection catalogs, we constructed a spec-confirmed LAE sample with 248 LAEs that have two or more relatively isolated components in a circular aperture of 2$\arcsec$ in diameter. 

We have studied the properties of the individual LAE components using the HST and JWST data. We estimated the photo-$z$ of all LAE components and found that $68.0\%$ of them are `real components' at the LAE redshifts. The remaining components are foreground (majority) and background (minority) objects. The fraction of `real component' decreases with the projected distance to the central `real component', from $\sim80\%$ at $0\farcs2-0\farcs4$ to $\sim30\%$ at $0\farcs8-1\farcs0$. 

For LAEs with only one `real component', which must be the Ly$\alpha$ emitting component, we found that they are young, low-mass, low dust-extinction star-forming galaxies, consistent with the previous result in the literature. In addition, all these Ly$\alpha$ emitting components have sSFR $>10^{-8.1} \, \rm yr^{-1}$, suggesting that they are starburst galaxies (partly due to the selection effect). For LAEs with two or more `real components', $90\%$ of their `real components' have sSFR $>10^{-8.1} \, \rm yr^{-1}$, suggesting that most of them likely emit Ly$\alpha$ photons. 

Most high-redshift LAEs appear as point-like sources in ground-based images, so much large aperture sizes are used for photometry in ground-based studies. We used an aperture of $2\arcsec$ in diameter and simulated ground-based images to perform photometry for the LAEs in our sample and repeated the above calculation of the physical properties without distinguishing between real LAE components and foreground/background contaminants. By comparing with the original measurements, we found that the usage of the large aperture size and the simulated ground-based images introduces a large bias to the magnitudes, which would largely affect the measurements of key properties such as stellar mass and SFR, etc. This finding suggests that multi-component LAEs should be taken into consideration seriously in future studies.

\begin{acknowledgments}

We thank the referee for comments that strengthened this manuscript.
We acknowledge support from the National Science Foundation of China (12225301). This work is based on observations taken by the CANDELS Multi-Cycle Treasury Program with the NASA/ESA HST, which is operated by the Association of Universities for Research in Astronomy, Inc., under NASA contract NAS5-26555.
This research made use of Photutils, an Astropy package for detection and photometry of astronomical sources \citep{larry_bradley_2024_13989456}.

\end{acknowledgments}

\vspace{5mm}
\facilities{HST (ACS), JWST (NIRCam)}

\software{Astropy \citep{2013A&A...558A..33A, 2018AJ....156..123A, 2022ApJ...935..167A}, 
          SExtractor \citep{1996A&AS..117..393B},
          Photutils \citep{larry_bradley_2024_13989456},
          PYPHER \citep{2016A&A...596A..63B},
          GALFIT \citep{2010AJ....139.2097P},
          EAZY \citep{2008ApJ...686.1503B},
          CIGALE \citep{2019A&A...622A.103B}
          }

\bibliography{ms}{}
\bibliographystyle{aasjournal}

\newpage

\appendix

\begin{figure*}[h]
\figurenum{A1}
\epsscale{0.92}
\plotone{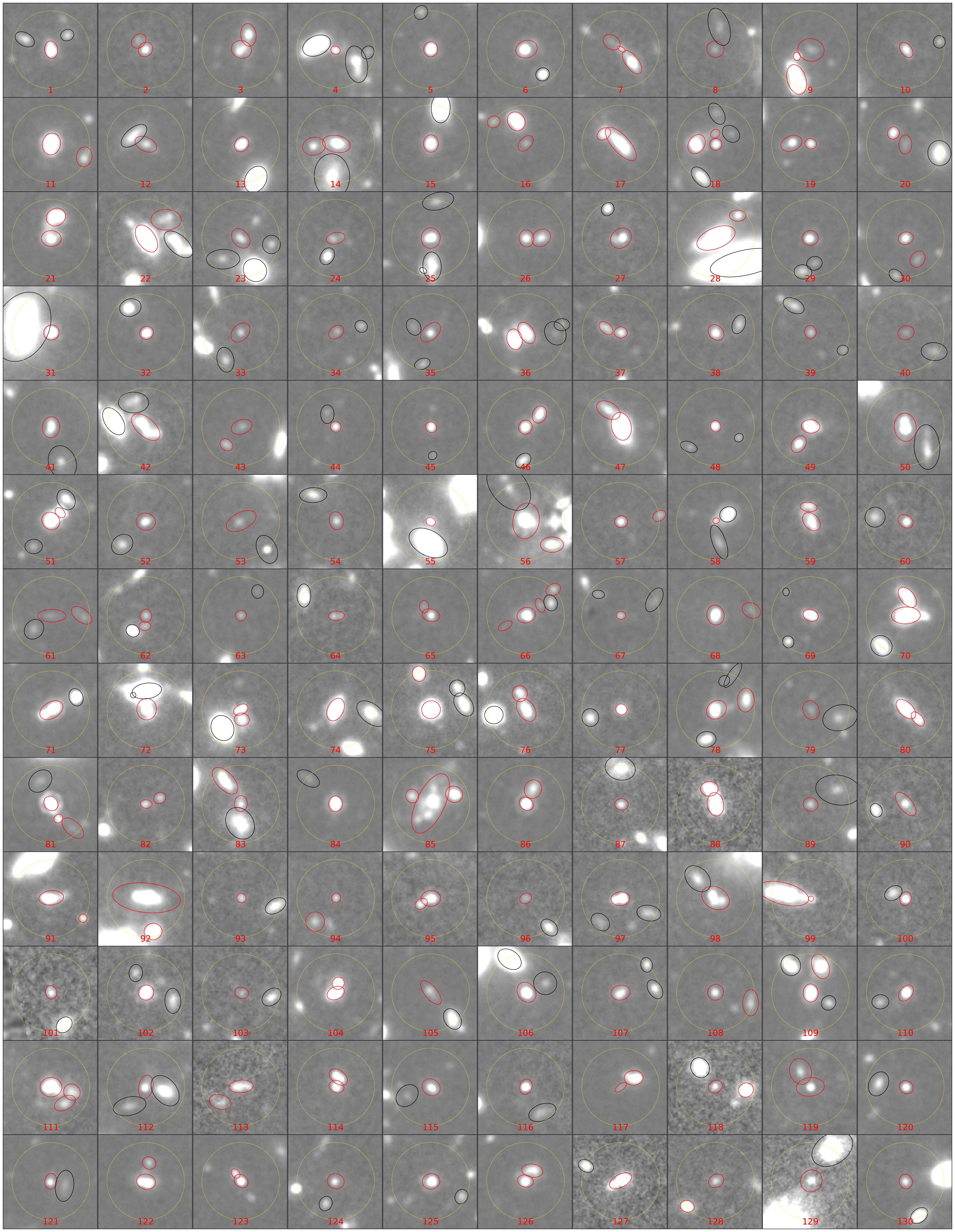}
\caption{Cutout images of all LAEs and components in the detection images. The size of cutout image is $2\farcs4 \times 2\farcs4$. The detection image in each field is made by inverse-variance-weighted combining PSF-homogenized images in the F200W, F277W, F356W, F410M and F444W bands. Red and black ellipses are the apertures used for the photometry of the components. Red ellipse refers to a `real component' while black ellipse refers to a foreground/background component. Yellow circles are circular apertures of $2\arcsec$ diameter centered on the central `real components' of the LAEs. Red numbers are the LAE-ID shown in Table \ref{tab:allinfotable}.
\label{fig:Allcomp1}}
\end{figure*}

\clearpage

\begin{figure*}[h]
\figurenum{A1}
\epsscale{0.92}
\plotone{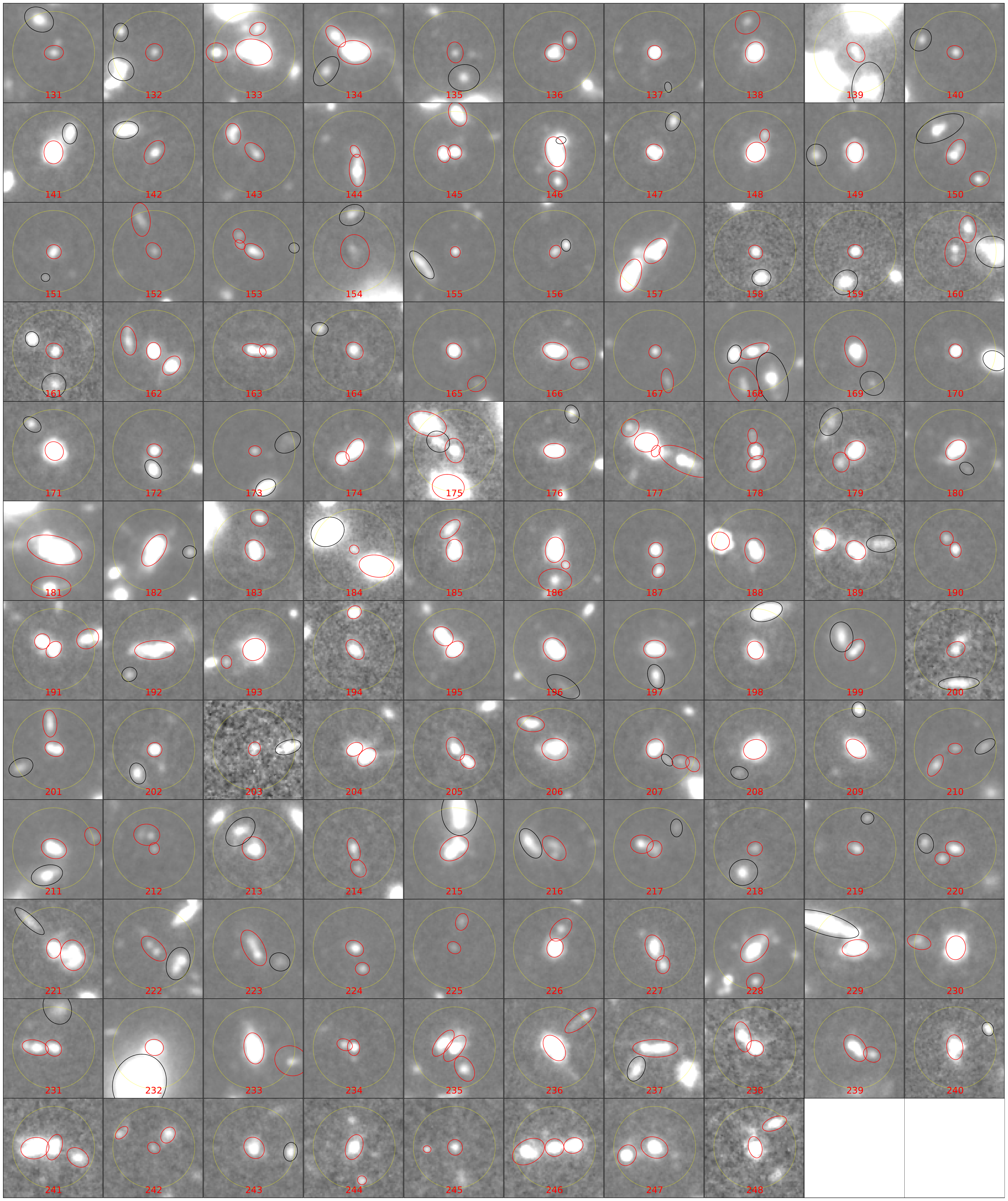}
\caption{(Continued.)
\label{fig:Allcomp2}}
\end{figure*}

\begin{deluxetable*}{cccccccccccc}
\tabletypesize{\scriptsize}
\tablewidth{0pt}
\tablenum{A1}
\tablecaption{Properties of the LAEs and components in the LAE sample \label{tab:allinfotable}}
\tablehead{
\colhead{LAE-ID} & \colhead{Reference} & \colhead{Field} & \colhead{Redshift} & \colhead{Comp-ID} & \colhead{R.A.} & \colhead{Decl.} & \colhead{Distance} & \colhead{Photo-$z$} & \colhead{$M_*$} &
\colhead{SFR} & \colhead{Age} \\
\colhead{} & \colhead{} & \colhead{} & \colhead{} & \colhead{} & \colhead{(J2000)} & \colhead{(J2000)} &
\colhead{(arcsec)} & \colhead{} & \colhead{($10^8 \, M_{\odot}$)} & \colhead{($M_{\odot} \, \rm yr^{-1}$)} & \colhead{($10^7\,$yr)}
}
\colnumbers
\startdata
1 & Kerutt+22 & GOODS-S & 2.942 & 1-1 & 53.133260 & $-27.786784$ & 0.00 & Yes & $2.4 \pm 1.2$ & $3.3 \pm 1.7$ & $22.6 \pm 24.4$\\
1 & Kerutt+22 & GOODS-S & 2.942 & 1-2 & 53.133130 & $-27.786679$ & 0.56 & No & ... & ... & ...\\
1 & Kerutt+22 & GOODS-S & 2.942 & 1-3 & 53.133473 & $-27.786709$ & 0.73 & No & ... & ... & ...\\
\hline
2 & Kerutt+22 & GOODS-S & 2.994 & 2-1 & 53.184429 & $-27.784443$ & 0.00 & Yes & $0.5 \pm 0.3$ & $2.1 \pm 0.6$ & $6.2 \pm 6.0$\\
2 & Kerutt+22 & GOODS-S & 2.994 & 2-2 & 53.184485 & $-27.784383$ & 0.28 & Yes & $0.2 \pm 0.1$ & $0.7 \pm 0.3$ & $5.6 \pm 5.0$\\
\hline
3 & Kerutt+22 & GOODS-S & 3.152 & 3-1 & 53.128321 & $-27.788707$ & 0.00 & Yes & $1.2 \pm 0.7$ & $3.2 \pm 1.4$ & $12.9 \pm 16.8$\\
3 & Kerutt+22 & GOODS-S & 3.152 & 3-2 & 53.128265 & $-27.788602$ & 0.42 & Yes & $2.9 \pm 1.4$ & $1.6 \pm 1.0$ & $49.1 \pm 37.0$\\
\hline
4 & Kerutt+22 & GOODS-S & 3.175 & 4-1 & 53.155229 & $-27.815267$ & 0.00 & Yes & $0.3 \pm 0.2$ & $1.2 \pm 0.4$ & $6.7 \pm 6.6$\\
4 & Kerutt+22 & GOODS-S & 3.175 & 4-2 & 53.155381 & $-27.815234$ & 0.50 & No & ... & ... & ...\\
4 & Kerutt+22 & GOODS-S & 3.175 & 4-3 & 53.155059 & $-27.815368$ & 0.65 & No & ... & ... & ...\\
4 & Kerutt+22 & GOODS-S & 3.175 & 4-4 & 53.154969 & $-27.815283$ & 0.83 & No & ... & ... & ...\\
\hline
5 & Kerutt+22 & GOODS-S & 3.300 & 5-1 & 53.178184 & $-27.790253$ & 0.00 & Yes & $1.9 \pm 1.1$ & $3.3 \pm 1.3$ & $16.9 \pm 17.7$\\
5 & Kerutt+22 & GOODS-S & 3.300 & 5-2 & 53.178264 & $-27.789990$ & 0.98 & No & ... & ... & ...\\
\hline
6 & Kerutt+22 & GOODS-S & 3.319 & 6-1 & 53.127519 & $-27.785046$ & 0.00 & Yes & $2.4 \pm 1.1$ & $3.4 \pm 1.9$ & $20.4 \pm 20.0$\\
6 & Kerutt+22 & GOODS-S & 3.319 & 6-2 & 53.127388 & $-27.785227$ & 0.77 & No & ... & ... & ...\\
\hline
7 & Kerutt+22 & GOODS-S & 3.361 & 7-1 & 53.146934 & $-27.790245$ & 0.00 & Yes & $0.1 \pm 0.1$ & $0.4 \pm 0.1$ & $9.2 \pm 7.2$\\
7 & Kerutt+22 & GOODS-S & 3.361 & 7-2 & 53.147013 & $-27.790195$ & 0.31 & Yes & $0.2 \pm 0.1$ & $0.6 \pm 0.2$ & $8.4 \pm 7.2$\\
7 & Kerutt+22 & GOODS-S & 3.361 & 7-3 & 53.146847 & $-27.790339$ & 0.44 & Yes & $5.1 \pm 1.9$ & $2.5 \pm 1.2$ & $55.9 \pm 33.7$\\
\hline
8 & Kerutt+22 & GOODS-S & 3.431 & 8-1 & 53.158068 & $-27.764901$ & 0.00 & Yes & $0.2 \pm 0.1$ & $0.7 \pm 0.3$ & $7.4 \pm 6.8$\\
8 & Kerutt+22 & GOODS-S & 3.431 & 8-2 & 53.158036 & $-27.764741$ & 0.59 & No & ... & ... & ...\\
\hline
9 & Kerutt+22 & GOODS-S & 3.433 & 9-1 & 53.169845 & $-27.768194$ & 0.00 & Yes & $0.4 \pm 0.2$ & $1.8 \pm 0.6$ & $5.7 \pm 6.2$\\
9 & Kerutt+22 & GOODS-S & 3.433 & 9-2 & 53.169959 & $-27.768240$ & 0.40 & Yes & $0.3 \pm 0.1$ & $1.7 \pm 0.5$ & $4.4 \pm 3.2$\\
9 & Kerutt+22 & GOODS-S & 3.433 & 9-3 & 53.169961 & $-27.768406$ & 0.85 & Yes & $12.2 \pm 4.6$ & $13.6 \pm 6.4$ & $22.5 \pm 17.8$\\
\hline
10 & Kerutt+22 & GOODS-S & 3.461 & 10-1 & 53.144192 & $-27.785200$ & 0.00 & Yes & $1.2 \pm 0.6$ & $1.7 \pm 0.9$ & $22.2 \pm 23.7$\\
10 & Kerutt+22 & GOODS-S & 3.461 & 10-2 & 53.143924 & $-27.785142$ & 0.88 & No & ... & ... & ...\\
\hline
11 & Kerutt+22 & GOODS-S & 3.468 & 11-1 & 53.141654 & $-27.805841$ & 0.00 & Yes & $9.7 \pm 5.1$ & $15.8 \pm 8.1$ & $19.2 \pm 21.6$\\
11 & Kerutt+22 & GOODS-S & 3.468 & 11-2 & 53.141389 & $-27.805933$ & 0.91 & Yes & $1.0 \pm 0.4$ & $2.5 \pm 1.2$ & $7.5 \pm 6.8$\\
\hline
12 & Kerutt+22 & GOODS-S & 3.497 & 12-1 & 53.173220 & $-27.767843$ & 0.00 & Yes & $0.7 \pm 0.4$ & $3.1 \pm 1.1$ & $6.6 \pm 8.3$\\
12 & Kerutt+22 & GOODS-S & 3.497 & 12-2 & 53.173316 & $-27.767783$ & 0.38 & No & ... & ... & ...\\
\hline
13 & Kerutt+22 & GOODS-S & 3.521 & 13-1 & 53.154104 & $-27.798849$ & 0.00 & Yes & $2.9 \pm 1.4$ & $2.9 \pm 1.6$ & $31.9 \pm 28.8$\\
13 & Kerutt+22 & GOODS-S & 3.521 & 13-2 & 53.153990 & $-27.799101$ & 0.98 & No & ... & ... & ...\\
\hline
14 & Kerutt+22 & GOODS-S & 3.556 & 14-1 & 53.157590 & $-27.764763$ & 0.00 & Yes & $4.4 \pm 1.9$ & $4.6 \pm 2.6$ & $27.6 \pm 24.0$\\
14 & Kerutt+22 & GOODS-S & 3.556 & 14-2 & 53.157771 & $-27.764779$ & 0.58 & Yes & $0.7 \pm 0.3$ & $2.3 \pm 0.8$ & $7.7 \pm 6.1$\\
14 & Kerutt+22 & GOODS-S & 3.556 & 14-3 & 53.157626 & $-27.764993$ & 0.84 & No & ... & ... & ...\\
\hline
15 & Kerutt+22 & GOODS-S & 3.598 & 15-1 & 53.153018 & $-27.806069$ & 0.00 & Yes & $2.4 \pm 1.2$ & $5.1 \pm 2.7$ & $13.8 \pm 14.7$\\
15 & Kerutt+22 & GOODS-S & 3.598 & 15-2 & 53.152939 & $-27.805815$ & 0.95 & No & ... & ... & ...\\
\hline
16 & Kerutt+22 & GOODS-S & 3.600 & 16-1 & 53.151710 & $-27.802801$ & 0.00 & Yes & $0.2 \pm 0.1$ & $1.0 \pm 0.4$ & $5.4 \pm 5.9$\\
16 & Kerutt+22 & GOODS-S & 3.600 & 16-2 & 53.151785 & $-27.802644$ & 0.62 & Yes & $7.1 \pm 2.9$ & $5.8 \pm 3.2$ & $35.4 \pm 30.0$\\
16 & Kerutt+22 & GOODS-S & 3.600 & 16-3 & 53.151968 & $-27.802647$ & 0.99 & Yes & $0.4 \pm 0.2$ & $0.5 \pm 0.3$ & $24.9 \pm 25.4$\\
\hline
17 & Kerutt+22 & GOODS-S & 3.600 & 17-1 & 53.160854 & $-27.801203$ & 0.00 & Yes & $8.6 \pm 4.1$ & $8.4 \pm 4.7$ & $30.7 \pm 28.2$\\
17 & Kerutt+22 & GOODS-S & 3.600 & 17-2 & 53.160992 & $-27.801132$ & 0.51 & Yes & $1.0 \pm 0.5$ & $2.5 \pm 1.2$ & $13.3 \pm 16.9$\\
\hline
18 & Kerutt+22 & GOODS-S & 3.666 & 18-1 & 53.178137 & $-27.774067$ & 0.00 & Yes & $1.6 \pm 0.6$ & $2.2 \pm 1.0$ & $13.8 \pm 17.2$\\
18 & Kerutt+22 & GOODS-S & 3.666 & 18-2 & 53.178142 & $-27.773991$ & 0.27 & Yes & $0.6 \pm 0.2$ & $0.3 \pm 0.3$ & $41.3 \pm 34.5$\\
18 & Kerutt+22 & GOODS-S & 3.666 & 18-3 & 53.178015 & $-27.773992$ & 0.48 & No & ... & ... & ...\\
18 & Kerutt+22 & GOODS-S & 3.666 & 18-4 & 53.178293 & $-27.774066$ & 0.50 & Yes & $2.3 \pm 1.2$ & $12.0 \pm 3.5$ & $4.6 \pm 3.6$\\
18 & Kerutt+22 & GOODS-S & 3.666 & 18-5 & 53.178129 & $-27.773847$ & 0.79 & No & ... & ... & ...\\
18 & Kerutt+22 & GOODS-S & 3.666 & 18-6 & 53.178257 & $-27.774301$ & 0.93 & No & ... & ... & ...\\
\hline
19 & Kerutt+22 & GOODS-S & 3.667 & 19-1 & 53.185239 & $-27.773518$ & 0.00 & Yes & $0.7 \pm 0.3$ & $2.7 \pm 1.0$ & $5.6 \pm 3.6$\\
19 & Kerutt+22 & GOODS-S & 3.667 & 19-2 & 53.185395 & $-27.773516$ & 0.50 & Yes & $1.2 \pm 0.6$ & $3.2 \pm 1.6$ & $11.5 \pm 14.5$\\
\hline
20 & Kerutt+22 & GOODS-S & 3.704 & 20-1 & 53.175484 & $-27.792518$ & 0.00 & Yes & $0.5 \pm 0.3$ & $0.7 \pm 0.5$ & $20.1 \pm 23.0$\\
20 & Kerutt+22 & GOODS-S & 3.704 & 20-2 & 53.175581 & $-27.792437$ & 0.43 & Yes & $1.4 \pm 0.7$ & $1.9 \pm 1.1$ & $28.2 \pm 29.9$\\
20 & Kerutt+22 & GOODS-S & 3.704 & 20-3 & 53.175208 & $-27.792580$ & 0.91 & No & ... & ... & ...
\enddata
\tablecomments{Col.(1): LAE ID that the component belongs to. Col.(2): Paper that discovered the LAE. Cols.(3): Field of the LAE. Cols.(4): LAE spec-$z$. Col.(5): Component ID. Cols.(6)-(7): R.A. and Decl. of the component. Col.(8): Projected distance to the `central component'. Col.(9): Whether the component photo-$z$ is consistent with the LAE spec-$z$ (`Yes') or not (`No'). Col.(10)-(12): Stellar mass, star formation rate, and age of the `real component'. `...' indicates that this is not a `real component'. The table only shows the first 20 LAEs and their components. A full table is available in the electronic version.}
\end{deluxetable*}

\end{document}